\documentclass[twocolumn,preprintnumbers,amsmath,amssymb,floatfix,superscriptaddress,nofootinbib]{revtex4}
\usepackage{amsmath,hyperref,amssymb}
\usepackage{graphicx}
\usepackage{dcolumn}
\usepackage{bm}
\usepackage{verbatim}
\usepackage{tabularx}
\usepackage{slashed}
\usepackage{cancel}
\usepackage{hyperref}
\usepackage{color}

\numberwithin{equation}{section}

\def\be{\begin{equation}}
\def\ee{\end{equation}}
\def\bea{\begin{eqnarray}}
\def\eea{\end{eqnarray}}
\def\ba#1\ea{\begin{align}#1\end{align}}
\def\bg#1\eg{\begin{gather}#1\end{gather}}
\def\bm#1\em{\begin{multline}#1\end{multline}}
\def\bmd#1\emd{\begin{multlined}#1\end{multlined}}

\renewcommand{\t}{\tilde}
\newcommand{\tr}{\text{tr}}

\newcommand{\SigmaN}{{\Sigma_\text{NFL}}}

\begin{document}

\title{Thermal effects in non-Fermi liquid superconductivity} 

\author{Jeremias Aguilera Damia}
\affiliation{\small \it Centro At\'omico Bariloche, CNEA and CONICET, Bariloche, R8402AGP, Argentina}
\author{Mario Sol\'is}
\affiliation{\small \it Centro At\'omico Bariloche, CNEA and CONICET, Bariloche, R8402AGP, Argentina}
\author{Gonzalo Torroba}
\affiliation{\small \it Centro At\'omico Bariloche, CNEA and CONICET, Bariloche, R8402AGP, Argentina}

\date{\today}

\begin{abstract}
We revisit the interplay between superconductivity and quantum criticality when thermal effects from virtual static bosons are included. These contributions, which arise from an effective theory compactified on the thermal circle, strongly affect field-theoretic predictions even at small temperatures. We argue that they are ubiquitous in a wide variety of models of non-Fermi liquid behavior, and generically produce a parametric suppression of superconducting instabilities. We apply these ideas to non-Fermi liquids in $d=2$ space dimensions, obtained by coupling a Fermi surface to a Landau-damped soft boson. Extending previous methods developed for $d=3-\epsilon$ dimensions, we determine the dynamics and phase diagram. It features a naked quantum critical point, separated by a continuous infinite order transition from a superconducting phase with strong non-Fermi liquid corrections. We also highlight the relevance of these effects for (numerical) experiments on non-Fermi liquids.
\end{abstract}

\maketitle

\tableofcontents

\section{Introduction}

Quantum criticality is thought to play a central role in the dynamics and superconductivity of strongly correlated electronic systems~\cite{Sachdev, RevModPhys.79.1015, keimer4673high, SHE2010S911, Zaanen2004, Eschrig2006, PhysRevLett.124.076801, Shibauchi2014, Mathur1998, Putzke2014, Dahm2009, LeTacon2011, Park2006, Chatterjee9346, Ramshaw317}. The proliferation of emergent gapless modes interacting with the electrons leads to many exciting phenomena, broadly characterized as ``non-Fermi liquid'' (NFL) behavior . Over the last decades, there has been fundamental progress, both experimental and theoretical, on the physics of strongly correlated materials.\footnote{See for instance \cite{Schofield1999, Coleman2005, SSLeeReview2018} and references therein.}  But the mechanisms behind high $T_c$ superconductivity and its connection with quantum criticality are not fully understood yet.

The interplay between quantum effects and superconductivity is in general quite nontrivial due to two competing tendencies of soft bosons: i) the destruction of coherent quasiparticles; ii) the enhancement of the superconducting pairing channel. Theoretically, it is possible to envision a rich landscape of possibilities ranging from an NFL state completely hidden under a superconducting dome~\cite{Polchinski1994,Altshuler1994,Metlitski}, to having a naked quantum critical point (QCP)~\cite{Raghu:2015sna,Wang:2016hir}.\footnote{The possibility of a naked QCP is experimentally very relevant and interesting~\cite{PhysRevLett.72.3262, RevModPhys.73.797, Gegenwart2008, Grigera2001, RevModPhys.79.1015, doiron2007quantum, yelland2008quantum}.} However, tractable field theory models usually require some expansion parameter (like Migdal's approximation \cite{migdal1958interaction, abrikosov2012methods}), and this has been an obstacle for comparing with real materials or with numerical experiments, which often lack such a parameter.

The goal of this work is to further develop the field theory framework for analyzing the interplay between quantum criticality and superconductivity. This continues previous lines of research \cite{Raghu:2015sna, Wang:2016hir, Wang:2017teb, Damia:2019bdx, Damia:2020yiu}. We incorporate both quantum and finite temperature effects, relevant for comparison with experiments. Previous studies at zero temperature $T=0$ in two spatial dimensions found that superconductivity is enhanced in NFLs when soft bosons come from order parameters, and that a stable naked QCP does not exist; on the other hand, superconductivity was found to be parametrically decreased or destroyed in the case of emergent $U(1)$ gauge fields~\cite{Metlitski, lee2007amperean,  wang2014pairing}. 

At finite temperature we uncover a much richer dynamics that builds on two key points. First, we identify a parameter `$N$' that measures the ratio between quantum corrections to the self-energy and corrections to the superconducting gap. As discussed in Sec.~\ref{sec:thermal}, many different NFLs have $N>1$, while $N=1$ in more standard Fermi liquids. Secondly, we demonstrate that the exchange of virtual bosons of zero Matsubara frequency (static modes) gives new contributions that tend to decrease the superconducting instability and can even lead to a naked QCP with critical pairing interactions. 

These virtual static contributions are a bit similar to impurity effects and have been traditionally neglected on the basis of Anderson's theorem \cite{anderson1959theory, millis1988inelastic, PhysRevB.78.220507, 2016PhRvL.117o7001W}. However, we argue that they do not cancel precisely when $N>1$. As a result, the dynamics is qualitatively different depending on $N=1$ versus $N>1$. Furthermore, $N$ does not need to be large in order to observe the non-cancellation of thermal effects from static modes. Even in NFLs where the emergent bosons are order parameters, we then find that superconductivity can be enhanced or diminished depending on the parameter $N$. The notorious differences between $N=1$ and $N>1$ dynamics provide nontrivial signatures that would be very interesting to target in future numerical experiments.

The main general lesson from our analysis will be that it is crucial to take into account effects from virtual zero frequency bosons at finite temperature. These static modes will be seen to play a major role in the interplay between quantum criticality and superconductivity in correlated electronic systems with $N>1$. Intuitively, they arise from an effective theory in one less dimension, obtained by compactifying on the thermal circle. For this reason, they are not governed by the scaling laws of the $T=0$ theory. We will argue that the resulting phase diagram at finite $T$ is very different from that based on naive $T=0$ expectations. An explicit analysis of the equations governing the fermion self-energy and superconducting gap require some specific information about the model (dispersion relations, Fermi surface geometry, etc.). We will do this for the well-known QCP with dynamical scaling $\omega^{2/3} \sim p$ that obtains from coupling a Fermi surface to an overdamped boson. But we would like to stress here that the basic result on the role of the static modes and $N>1$ is more general and should be analyzed in other models as well.

This paper is organized as follows. First in Sec.~\ref{sec:thermal} we present a more or less general discussion of thermal effects, the contribution of virtual static bosonic modes and the role of the parameter $N$. While concrete calculations require more explicit model-dependent assumptions (the subject of the following sections), the goal here is to argue for the generality of such effects in NFLs. In Sec.~\ref{sec:model} we briefly review the more specific theory that we will study, obtained by coupling a Fermi surface to Landau damped bosons. Our main results are in Sec.~\ref{sec:finiteT}, where we study the superconducting instability at linearized level and its critical temperature, obtaining the phase diagram and exhibiting the key effects from $N$ and the thermal static modes. The normal state is found to be stable for $N> N_{cr} \sim 8$; a superconducting instability develops via an infinite-order transition as $N \to N_{cr}$, and the ground state is a superconductor for $N< N_{cr}$. This phase exhibits strong NFL and thermal effects, most notably a parametric suppression of the critical temperature compared to the physical gap. In Sec.~\ref{sc:non-linear} we explore the gap equation at nonlinear level, looking for solutions that may have been missed in the linearized treatment. We indeed find a different type of solution that is predominantly sourced by the first few Matsubara modes of the gap. However, we argue that the above thermal effects imply that this solution is never energetically preferred. Therefore our results for the dynamics and phase diagram appear consistent also at the nonlinear level. We also briefly compare this situation with the related (but eventually quite different) scenario of the ``first Matsubara law'' analyzed in recent works~\cite{2016PhRvL.117o7001W, 2019PhRvB..99n4512W, 2019arXiv191201797C, Abanov_2020, Wu_2020}. 
Finally, we summarize the main conclusions and future directions in Sec.~\ref{sec:concl}.  Two appendices contain somewhat more technical results that are used in the main text.

\section{Thermal effects on fermion self-energy and gap }\label{sec:thermal}

Non-Fermi liquid behavior arises generically from interactions between a Fermi surface and soft bosons. The same interaction that produces interesting quantum physics simultaneously leads to strong thermal corrections. Our goal in this section is to present a discussion of finite temperature effects at a general level, making manifest that they will be ubiquitous for a broad class of models. In the following sections we will perform a detailed analysis for a class of NFLs with Landau damped bosons.

Consider a coupling of the Yukawa form $H_{int}=g \phi \psi^\dag \psi$, and denote the boson propagator by $D(q, \Omega)$, with $q$ the momentum and $\Omega$ the frequency of the boson. At one and higher loops, virtual bosons lead to a fermion self-energy $\Sigma$ and, possibly, to the formation of a superconducting gap $\Delta$. Generically there will also be vertex corrections, but we will neglect them here.\footnote{In concrete models there can be some Migdal-type expansion that justifies this. This will be the case in the model below, using large $N$.} In order to exhibit the thermal corrections in their simplest form, it is sufficient to keep only the linear effects from the gap. This is also the physically relevant regime for studying the putative superconducting transition; in Sec.~\ref{sc:non-linear} we will analyze nonlinear effects from the gap. 

Given these simplifications, the fermion dynamics is determined by two self-consistent Schwinger-Dyson equations. The first fixes the self-energy,
\ba\label{eq:sec2Sigma}
i\Sigma(p,\omega_n&)= -g^2 T \sum_m\int\frac{d^dq}{(2\pi)^d}D(p-q,\omega_n-\omega_m) \nonumber\\
& \times \frac{1}{i \omega_m + i \Sigma(q, \omega_m) - \varepsilon_q}
\ea
where $\varepsilon_q$ is the classical fermion dispersion relation, $n, m$ are discrete Matsubara indices and $d$ is the number of spatial dimensions. The second self-consistent equation, also known as the Eliashberg equation, determines the gap
\ba\label{eq:sec2Delta}
\tilde \Delta(p,\omega_n) &= \frac{g^2}{N} T \sum_m \int \frac{d^dq}{(2\pi)^d}D(p-q,\omega_n-\omega_m)\nonumber\\
&\times  \frac{\tilde \Delta(q,\omega_m)}{\left(\omega_m + \Sigma(q, \omega_m) \right)^2+\varepsilon_{q}^2}\,. 
\ea
Here $\t \Delta$ is the pairing vertex, that appears directly in the Hamiltonian, $H \supset \t \Delta \psi_{p} \psi_{-p}+ c.c.$; it is related to the physical gap by
\be\label{eq:physDelta}
\tilde \Delta(p,\omega_n) = \left( 1 + \frac{\Sigma(p, \omega_n)}{\omega_n} \right) \Delta(p,\omega_n)\,.
\ee

We have introduced a parameter $N>1$ that distinguishes the strength of corrections to the self-energy and the gap. Let us explain our reasons and motivations behind this. In the simplest cases of BCS or phonon superconductivity, $N=1$, but $N>1$ is motivated by non-Fermi liquid physics. Indeed, one of the original examples for this is in color superconductivity, where $N=3$ from the $SU(3)$ gauge interaction of chromodynamics~\cite{Son:1998uk}. Strongly correlated electronic systems can also have matrix-type order parameters that give $N>1$, as in antiferromagnetic type materials~\cite{2012RvMP...84.1383S}. $N$ can also be related to unconventional superconductivity with nonzero angular momentum. 
More recently, new materials have been constructed that display approximate global symmetries that  translate to $N>1$. This includes the valley symmetry in graphene and could also be relevant for the exciting discoveries in twisted bilayer graphene~\cite{2018Natur.556...43C, cao2018magic, cao2020strange, po2018origin, wang2020topological, baskaran2018theory, roy2019unconventional, isobe2018unconventional, dodaro2018phases, you2019superconductivity}. Another motivation is to have QFT predictions to compare with future numerical experiments on the lattice. This is now especially relevant given recent progress in Monte Carlo methods~\cite{Schattner2016, 2017PNAS..114.4905L, Berg2019, liu2019itinerant, klein2020normal, Xu:2020tvb}.  One of our main results will be that $N>1$ leads to thermal effects that are qualitatively different from $N=1$, and this provides nontrivial signatures to look for numerically.

In the euclidean formalism, finite temperature contributions are represented by the Matsubara sums in (\ref{eq:sec2Sigma}) and (\ref{eq:sec2Delta}), with $\omega_n= \pi T(n+1/2)$. Small temperature means $n$ large, and in this limit one expects to recover the $T=0$ results plus corrections. However, the terms with $m=n$ in the sum can lead to large departures from this. They come from exchange of static bosons with $\Omega_n=0$. Such contributions can be analyzed with an effective theory in one less space-time dimension, obtained by compactifying on the thermal circle. As a result, static effects do not need to respect the scaling laws of the $T=0$ theory, and indeed can lead to large violations of these. Our goal is to determine how these thermal contributions affect superconductivity in non-Fermi liquids.

Virtual static bosons resemble impurities, and have often been neglected in previous works on the basis of Anderson's theorem \cite{anderson1959theory, millis1988inelastic, PhysRevB.78.220507, 2016PhRvL.117o7001W}. We will review that the reason for this is that they can be rescaled away when $N=1$. Crucially, however, this cancellation will be seen to fail for $N>1$, see Sec.~\ref{subsec:anderson} below. This is why thermal physics will be very different. Non-cancellation of thermal static effects will modify the behavior of the gap $\Delta$ and the superconducting transition $T_c$, as well as generically producing a parametric difference $T_c \ll \Delta$. This is potentially relevant for unconventional superconductors with such a hierarchy of scales~\cite{2012JPSJ...81a1006Y, 2016NatCo...712843K}.

The way to study and solve (\ref{eq:sec2Sigma}) and (\ref{eq:sec2Delta}) is somewhat model-dependent. In this work we will focus on a spherical Fermi surface and the well-known quantum critical scaling for the fermion $\Sigma(\omega) \sim \omega^{2/3}$. The methods we develop may also be applied to other dynamical exponents and/or geometries, and we hope to address these in future work. Despite this model dependence, let us make some remarks that should apply more generally to non-Fermi liquids. 

The first is that the non-cancellation of thermal effects is expected when $N \neq 1$.  This does not require $N \gg 1$, but works for any $N>1$. From this point of view, $N=1$ may be a choice that is non-generic for the dynamics of NFLs and their interplay with superconductivity. The second point to stress is that once thermal effects from static boson exchange contribute, they are expected to tend to diminish the superconducting gap, and correspondingly increase the NFL region above the superconducting dome. The reason is that, as we discussed before, we can think of them as arising from an effective theory in one less dimension, and decreasing the dimensionality tends to disorder the ground state.\footnote{The prototypical example is the Coleman-Mermin-Wagner theorem that implies no symmetry breaking in two space-time dimensions \cite{mermin1966absence, hohenberg1967existence, coleman1973there}. }  We will see explicitly how this comes about.
The parameter $N$ could then be relevant for explaining why certain NFLs have an enhanced superconducting temperature $T_c$, while in others $T_c$ is decreased. It would be interesting to revisit models of strange metals in light of the results in this work.

Finally, let us stress that we are considering a different situation from that of a series of recent works on the interplay between superconductivity and quantum criticality~\cite{2016PhRvL.117o7001W, 2019PhRvB..99n4512W, 2019arXiv191201797C, Abanov_2020, Wu_2020}. Those references analyze equations similar to (\ref{eq:sec2Sigma}) and (\ref{eq:sec2Delta}) but removing $m = n $ terms. Their focus is on $N=1$ where $m=n$ terms can be rescaled away, and the parameter $N$ in these works is a way of dialing the relative strength of quantum critical and superconducting contributions. In contrast, in our work $N$ is physical as we motivated before, and thermal effects from virtual static bosons will turn out to play a fundamental role. This is the reason why our results are different from those works; a more detailed comparison will be presented in Sec.~\ref{subsec:firstM}.

\section{Model and dynamics at zero temperature }\label{sec:model}

The rest of the paper is devoted to analyzing the interplay between quantum criticality and superconductivity at finite temperature in a concrete NFL model. We will demonstrate the failure of Anderson's cancellation of static effects for $N>1$, and we will obtain the phase diagram that is qualitatively changed due to thermal effects. 
In this section we review the model and study its dynamics at $T=0$. We first discuss the two-dimensional quantum critical point of~\cite{Damia:2019bdx}, based on a Fermi surface interacting with soft bosons via a Yukawa coupling. In the second step we add BCS 4-Fermi couplings and use the renormalization group and Schwinger-Dyson equations to study their effects on the fixed point.

\subsection{Non-Fermi liquid fixed point}\label{subsec:nfl1}

We consider a model of non-Fermi liquids where a Fermi surface  (taken to be spherical for simplicity) of fermions $\psi$ is coupled to a Landau-damped massless scalar $\phi$ with $z_b=3$ dynamical exponent. Following previous works~\cite{Fitzpatrickone, Mahajan2013, FKKRtwo, Torroba:2014gqa}, $\psi^i$ is promoted to an $N$-component field, while $\phi^i_j$ is an $N \times N$ matrix. 
This parameter $N>1$ will lead to the non-cancellation of thermal static effects.

We start from the following low energy effective action that respects the $SU(N)$ symmetry:
\be\label{eq:S00}
S=S_f+S_b+S_Y
\ee 
with
\bea\label{eq:S0}
S_f &=&- \int_{\omega, p} \psi^\dag_i ( i \omega - \varepsilon_p) \psi^i \nonumber\\
S_b&=& \frac{1}{2} \int_{\Omega, q} \phi_i^j \left(q^2 + M_D^2 \frac{|\Omega|}{q}\right) \phi_j^i \\
S_Y&=&\frac{g}{\sqrt{N}}   \int_{\omega, p}  \int_{\Omega, q} \phi_j^i(\Omega, q) \psi_i^\dag(\omega, p) \psi^j(\omega-\Omega, p-q) \nonumber\,.
\eea 
Here $M_D$ is the Landau damping mass scale, and we take it as the UV cutoff in our effective description. The bare mass for the boson is tuned to zero to approach the quantum critical point. Given this tuning, the Yukawa interaction with strength $g$ is the most relevant one consistent with the $SU(N)$ symmetry. Except for the marginal BCS scattering that we will introduce shortly, other interactions turn out to be irrelevant at the fixed point.

The spherical dispersion relation
\be
\varepsilon_p = \frac{p^2}{2m} - \mu_F\,,
\ee
gives a Fermi surface of radius $k_F=\sqrt{2m \mu_F}$. Since we are interested in the low energy/momenta dynamics, it will be sufficient to linearize
\be
\varepsilon_p \approx v p_\perp\;,\;\vec p \equiv \hat n (k_F + p_\perp)
\ee
with $\hat n$ a unit vector on the Fermi surface. This is the spherical RG of~\cite{Shankar}.

It was recently shown in~\cite{Damia:2019bdx} that, in the large $N$ limit, (\ref{eq:S00}) leads to a controlled quantum critical point with non-Fermi liquid behavior. Below the dynamical scale
\be
\Lambda = \frac{g^6}{(2\pi v \sqrt{3})^3 M_D^2}\,,
\ee 
the theory flows to a fixed point with fermion self-energy
\be\label{eq:SigmaNFL}
\SigmaN(\omega) = \Lambda^{1/3}{\rm sgn}(\omega) |\omega|^{2/3}\,,
\ee  
that gives a $z_f=3/2$ dynamical exponent. All other effects at the fixed point (including corrections to the boson 2-point function) are suppressed by $1/N$. We stress that we start with a Landau-damped $z=3$ boson, and that's why we take $M_D$ as the UV cutoff; this is required in order to avoid the large $N$ problems found in~\cite{Lee2009}. By itself, the low energy dynamics also generates a Landau-damping contribution, but it is suppressed by $1/N$.

Our goal in what follows is to study the competition between non-Fermi liquid behavior and the superconducting instability in this setup, first at zero temperature and then taking into account thermal effects.

\subsection{Incorporating the BCS interaction}

Let us now include the 4-Fermi interaction in the BCS channel. We briefly discuss both the renormalization group approach, following~\cite{Shankar, Polchinski, Son:1998uk, Metlitski, Raghu:2015sna}, and the Eliashberg method.
Boson exchange in (\ref{eq:S00}) leads to a non-local 4-Fermi interaction. The idea of the RG, first explained in~\cite{Son:1998uk}, is that part of this non-local term becomes local when integrating over momentum or frequency shells, and hence contributes to the BCS beta function. This approach was generalized in~\cite{Raghu:2015sna} to incorporate the fermion anomalous dimensions, and that is the version we will use here.

It is convenient to normalize the BCS coupling as follows,
\be
H_\text{BCS}=-\frac{v }{4k_FN}\lambda \,(\psi^\dag \psi \psi^\dag \psi)
\ee
where the fermion's momenta (not shown here) are on the BCS channel~\cite{Shankar, Polchinski}. The prefactor $v/(4 k_F)$ is chosen to simplify formulas, and the $1/N$ reflects the non-planar character of the BCS scattering. 
There are three contributions to the RG beta function of the BCS coupling $\lambda$~\cite{Raghu:2015sna}. Let us analyze them in turn.

First, tree-level boson exchange contributes a source term proportional to $g^2$ that accounts for the boson-mediated interaction that becomes local~\cite{Son:1998uk}.  Integrating out the boson generates an effective non-local interaction
\be
S_{int}= - \frac{g^2}{2N} \int _{q,p,p'}\,D(q)\,\psi_i^\dag(p) \psi^j(p-q)\,\psi_j^\dag(p') \psi^i(p'+q)\,.
\ee
This interaction is attractive, so it will be sufficient to consider $s$-wave pairing.\footnote{The Landau-damped boson leads to nontrivial dependence for higher angular momenta, see e.g.~\cite{2013PhRvB..88d5127C, 2014AnPhy.351..727W}.} Using this and the $z_b=3$ scaling of the boson\footnote{This allows to factorize the integral into perpendicular and tangential momenta to the Fermi surface.}
leads to a frequency-dependent pairing
\be\label{eq:nonlocal}
\frac{g^2}{2N}\,\int\frac{dq_\parallel}{2\pi}\,D(\Omega, q_\parallel) = \frac{1}{N}\,\frac{g^2}{3 \sqrt{3}}\,\frac{1}{(M_D^2 |\Omega|)^{1/3}}
\ee
where $q_\parallel$ here is the bosonic momentum orthogonal to the local $\vec k_F$. The tree-level contribution to the beta function of the BCS coupling is obtained by deriving (\ref{eq:nonlocal}) with respect to $\Omega$.

The next contribution comes about because in a non-Fermi liquid the fermions acquire a positive anomalous dimension, and this makes the BCS coupling irrelevant~\cite{Raghu:2015sna}. At large $N$, the fermion anomalous dimension is $\gamma=1/6$ (defined as changing the classical frequency term by $\omega \to \omega^{1-2\gamma}$). By itself, this is a large effect at the fixed point, giving an irrelevant scaling dimension $[\lambda]=-1/3$. If it were for this contribution alone, we would have $\lambda \to 0$.
The last contribution comes from the one loop fermion bubble~\cite{Shankar}, and this is the standard contribution in Fermi liquids. 

The resulting beta function at the QCP evaluates to
\be\label{eq:BCSRG}
\frac{d\lambda}{d\log\mu}= -\frac{8\pi}{9}+\frac{ \lambda}{3} -\frac{\lambda^2}{4\pi N} \,.
\ee
This encodes mathematically the competition between non-Fermi liquid and pairing effects in the beta function: the tree-level exchange and one loop fermion bubble push to make $\lambda$ relevant, while the NFL anomalous dimension tries to make the BCS coupling irrelevant.

\begin{figure}[h]
  \centering
  \includegraphics[width=1.\hsize]{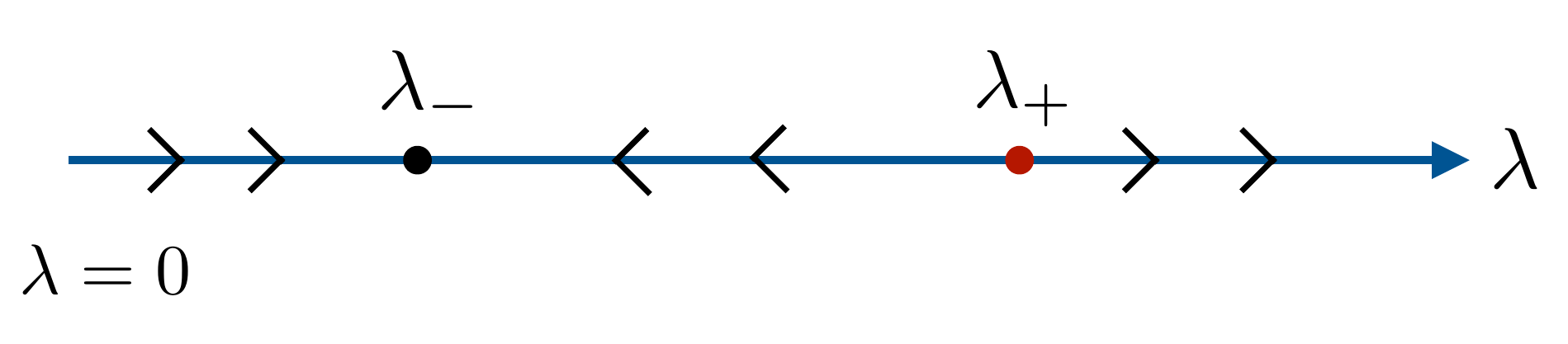}
  \caption{Schematic RG flow of the BCS coupling between UV and IR fixed points.}
  \label{fig:lambda-flow}
\end{figure}

A general consequence of beta functions like (\ref{eq:BCSRG}) is that they support UV and IR fixed points and give rise to Berezinskii-Kosterlitz-Thouless (BKT) scaling when the fixed points annihilate~\cite{Kaplan:2009kr}. In our case, the fixed points are at
\be\label{eq:lambdapm}
\lambda_\pm = \frac{2\pi}{3} N\left( 1 \pm \sqrt{1-8/N}\right)\,,
\ee
valid for $N>8$. From the slope of the beta function, $\lambda_-$ is attractive while $\lambda_+$ is repulsive. The resulting RG flow is illustrated in Fig.~\ref{fig:lambda-flow}.

The stable fixed point at
\be
\lambda^* = \frac{2\pi}{3} N\left( 1 - \sqrt{1-8/N}\right)\approx \frac{8\pi}{3}\;\;(\text{for}\;N \gg 1)
\ee
describes a quantum critical point for the BCS coupling. This was one of the main findings of~\cite{Raghu:2015sna}, a metallic state with critical BCS interactions. The same mechanism discovered in that work, which used an epsilon expansion around $d=3$, also applies here in $d=2$ by virtue of the large $N$ limit. While this is outside the scope of this paper, it would be interesting to understand the phenomenological implications of critical pairing interactions in the physical dimension $d=2$.

Ref.~\cite{Raghu:2015sna} also identified the transition between NFL and superconducting states as an infinite order BKT transition driven by fixed point annihilation. We see the same applies here: as $N \to 8$, the UV and IR fixed points merge and for $N<8$ they go off into the complex plane. In the absence of a stable fixed point, the RG flow of Fig.\ref{fig:lambda-flow} then proceeds to the $\lambda \to \infty $ region; this growth in the pairing attraction signals a superconducting instability, and we will soon analyze this from the point of view of the gap equation. We will see that  the value $N_{cr}=8$ receives small corrections from non-local effects that are not captured by the RG. Furthermore such values may be outside the regime of validity of the large $N$ approximation that we have used.\footnote{Nevertheless, experience from similar matrix large $N$ limits in QCD suggests that even values like $N \sim 3$ may still be qualitatively correct~\cite{Witten:1979kh}.}  It seems plausible that the transition between the normal and superconducting states in the full theory will still occur via BKT scaling, but it would be interesting to analyze the leading $1/N$ corrections and their effect on the fixed-point annihilation picture. 

\subsection{Eliashberg equation and BKT transition}\label{subsec:SD}

An alternative approach is to allow for a superconducting gap and derive a self-consistent equation that it should obey.
The pairing vertex appears in the Hamiltonian as
$H \supset \t \Delta \psi_{p} \psi_{-p}+c.c.$, and is related to the physical gap $\Delta$ via (\ref{eq:physDelta}) above. A very useful simplification near QCPs like the one we are considering is that both the self-energy and the gap depend predominantly on frequency; this restricts integral and differential equations to just one variable. The self-consistent Schwinger-Dyson equation at $T=0$ is
\be\label{eq:tDelta-nonlinear}
\t \Delta(p)= \frac{g^2}{N}\,\int \frac{d^3q}{(2\pi)^3}\,\frac{D(p-q)\,\t \Delta(q)}{|q_0+\Sigma(q_0)|^2+|\t \Delta(q)|^2+\varepsilon_q^2}\,.
\ee
If the system superconducts, the frequency dependence will stabilize below the physical gap scale, defined as
\be\label{eq:gap-scale}
\Delta_0 \equiv \Delta(\omega=\Delta_0)\,.
\ee
As is standard with gap equations, we will approximate this by linearizing the integrand in (\ref{eq:tDelta-nonlinear}), while restricting the frequency integral to $|\omega|> \Delta_0$. This value $\Delta_0$  will have to be determined self-consistently.

As in (\ref{eq:nonlocal}), 
the $z_b=3$ scaling implies that the $q_\parallel$ integral affects only the boson propagator, while $q_\perp$ appears only in the fermion propagator. Performing the integrals obtains
\be\label{eq:zeroTSDE}
\t \Delta(\omega)=  \frac{\Lambda^{1/3}}{3N}\,\int_{|\omega'|>\Delta_0}d\omega' \frac{1}{|\omega-\omega'|^{1/3}} \frac{\t \Delta(\omega')}{|\omega' +\Sigma(\omega')|}\,.
\ee
This is the familiar Eliashberg equation for the gap, here evaluated with retardation effects from the soft $z_b=3$ boson. We focus on a gap that is even in frequency. Then the linearized gap equation rewrites to
\be
\t \Delta(\omega)= \frac{1}{2N}\int_{\Delta_0}^{\Lambda_0} d\omega'\,u(\omega,\omega')\frac{\t \Delta(\omega')}{A(\omega')}\label{eq:zeroTSDE1}
\ee
with $\Lambda_0$ a UV frequency cutoff, $A(\omega)=\omega+\Lambda^{1/3}|\omega|^{2/3}$ and 
\be
u(\omega,\omega') = \frac{2\Lambda^{1/3}}{3}\left(\frac{1}{|\omega-\omega'|^{1/3}}+\frac{1}{|\omega+\omega'|^{1/3}}\right)\,.
\label{eq:int-kernel}
\ee

As shown in~\cite{Wang:2016hir} in the analog problem in $d=3-\epsilon$ dimensions, the RG result is recovered in a local approximation where frequency mixing in (\ref{eq:zeroTSDE}) is small. We will find that this is also a very good approximation in $d=2$, up to small corrections that will be determined via numerics and will be incorporated below in Sec.~\ref{sec:finiteT}.

By means of a local approximation $u(\omega,\omega')\approx 2 u(\omega)$ ($u(\omega,\omega')\approx2 u(\omega')$) for $\omega>\omega'$ ($\omega<\omega'$), \eqref{eq:zeroTSDE1} can be mapped to a differential equation of the form
\be
\frac{d}{d\omega}\left(\frac{\t \Delta'(\omega)}{u'(\omega)}\right)=\frac{\t \Delta(\omega)}{N A(\omega)}
\label{eq:zeroTdiff}
\ee 
together with the UV and IR boundary conditions inherited from the integral equation,
\be\label{eq:bc}
\frac{d}{d\omega}\left(\frac{\t \Delta(\omega)}{u(\omega)} \right)\Big|_{\omega=\Lambda_0}=0\;,\;\tilde\Delta'(\omega)\Big|_{\omega=\Delta_0}=0\,.
\ee
The change of variables
\be
\lambda(\omega) = \frac{8\pi}{3}\, \frac{u'(\omega)}{u(\omega)}\, \frac{\t \Delta(\omega)}{\t \Delta'(\omega)}
\label{eq:maptoRG}
\ee
transforms the differential problem into the RG beta function, thus establishing their equivalence. 

Let us now analyze the transition between quantum criticality and superconductivity using the gap equation. Following~\cite{Wang:2016hir} (similar calculations have also appeared more recently in~\cite{2019arXiv191201797C}), we change variables to
\be
\omega= \Lambda\, e^{-3x}\;,\;g_1=\frac{2}{N}
\label{eq:x}
\ee 
and the differential equation \eqref{eq:zeroTdiff} reads
\be
(e^{-x}+1)(\t \Delta''(x)-\t \Delta'(x))+g_1\t \Delta(x)=0\,.
\label{eq:x-diffT0}
\ee
Up to an overall scale that is not fixed by the linearized analysis, the solution is
\ba\label{eq:hyper}
\t\Delta(x)&= e^x \,_2F_1\left(\frac{1}{2}-\frac{1}{2}\sqrt{1-4g_1},\frac{1}{2}+\frac{1}{2}\sqrt{1-4g_1},2,-e^x\right) \nonumber\\
&+ C_\Lambda\text{MeijerG}\left(\left\lbrace \lbrace\rbrace,\left\lbrace\frac{3}{2}-\frac{1}{2}\sqrt{1-4g_1},\frac{3}{2} \right. \right. \right. \nonumber\\
&+\left. \left. \left. \frac{1}{2}\sqrt{1-4g_1}\right\rbrace\right\rbrace,\left\lbrace \lbrace 0,1\rbrace,\lbrace\rbrace\right\rbrace, -e^x\right)\,,
\ea
with $C_\Lambda$ fixed by the UV boundary condition in (\ref{eq:bc})~\cite{Wang:2016hir}.

To develop intuition, let us approximate this
in both the high and low frequency regimes, that is, $x<0$ and $x>0$ respectively. The piecewise solution reads
\be
\t \Delta(x) \approx \left\lbrace 
\begin{array}{lcc}
e^{\frac x 2}J_1(2\sqrt{g_1}e^{\frac x 2}) & , & -\infty < x < 0 \\
C_1 e^{\frac x 2} \cos\left(\sqrt{g_1-\frac{1}{4}}x+\phi \right) & , & 0<x<x_0  \\
D_1 & , & x>x_0
\end{array}
\right.
\label{eq:app-solT0}
\ee
where $x_0$ is related to $\Delta_0$ by \eqref{eq:x}, namely
\be\label{eq:x0Delta0}
\Delta_0= \Lambda e^{-3 x_0}\,.
\ee
The integration constants $C_1$ and $\phi$ are implicit functions of $g_1$ fixed by gluing both solutions at $x=0$ or by comparing with (\ref{eq:hyper}).

The physical gap scale $\Delta_0$ is determined self-consistently by the IR boundary condition $\tilde\Delta'(x_0)=0$. Crucially, this has solution only for $g_1\geq1/4$, so $N_{cr}=8$ for the transition between the normal and superconducting states. We note here that the integral equation actually gives $N_{cr} \approx 8.3$, and we use this value in what follows; the difference with the RG or local approximation result $N_{cr}=8$ is discussed in more detail in Sec.~\ref{sec:finiteT}.
 For $g_1\geq1/4$, we get
\be
x_0 = \frac{\pi-2\phi(g_1)}{\sqrt{4g_1-1}}
\ee 
where $\phi(g_1)\approx -\pi/2$ for $g_1 \sim 1/4$. Plugging back into (\ref{eq:x0Delta0}), we obtain the physical gap
\be
\Delta_0 \approx \Lambda \exp\left(-\frac{6\pi}{\sqrt{\frac{N_{cr}}{N}-1}}\right)\,.
\label{eq:cr-Delta0}
\ee

The physical gap scale vanishes non-analytically as $N\to N_{cr}$, exhibiting BKT scaling. This agrees qualitatively with the RG intuition of fixed point annihilation, and extends the $d=3-\epsilon$ results of~\cite{Raghu:2015sna, Wang:2016hir, Wang:2017teb} to $d=2$. BKT behavior was also recently observed by~\cite{Abanov_2020} in the $\gamma$-model~\cite{Moon2010} and by~\cite{PhysRevLett.124.017002} in the Yukawa SYK model. Away from $N \approx N_{cr}$, the physical gap can be obtained by numerically solving the equation imposed by the IR boundary condition for the full solution (\ref{eq:hyper}). This shows a very good agreement with the estimate \eqref{eq:cr-Delta0}.

\section{NFL superconductivity at finite temperature}\label{sec:finiteT}

Our main goal in this work is to understand the interplay between superconductivity and quantum criticality in 2d non-Fermi liquids at finite temperature. At large $N$ vertex corrections can be neglected, so we need to solve the self-consistent Schwinger-Dyson equations for the boson self-energy $\Pi$, fermion self-energy $\Sigma$, and fermion gap/pairing vertex $\t \Delta$ (see Appendix \ref{ap:SD})

\begin{widetext}
\bea\label{eq:TSD}
\Pi(\Omega_m ,q) &=& \frac{g^2}{N} T \sum_n\,\int \frac{d^2 p}{(2\pi)^2}\,\frac{i A(\omega_n) +\varepsilon_p}{A(\omega_n)^2+|\tilde \Delta(\omega_n)|^2+\varepsilon_p^2}\frac{i A(\omega_n+\Omega_m) +\varepsilon_{p+q}}{A(\omega_n+\Omega_m)^2+|\tilde \Delta(\omega_n+\Omega_m)|^2+\varepsilon_{p+q}^2} \nonumber\\
i \Sigma(\omega_n) &=& g^2 T \sum_m \int \frac{d^2q}{(2\pi)^2} \frac{1}{q^2+M_D^2 \frac{|\Omega|}{q}+\Pi(\omega_m-\omega_n,q)}\,\frac{i A(\omega_m)+\varepsilon_{p+q}}{A(\omega_m)^2+|\tilde \Delta(\omega_m)|^2+\varepsilon_{p+q}^2} \\
\tilde \Delta(\omega_n) &=& \frac{g^2}{N} T \sum_m \int \frac{d^2q}{(2\pi)^2} \frac{1}{q^2+M_D^2 \frac{|\Omega|}{q}+\Pi(\omega_m-\omega_n,q)}\,\frac{\tilde \Delta(\omega_m)}{A(\omega_m)^2+|\tilde \Delta(\omega_m)|^2+\varepsilon_{p+q}^2}\,. \nonumber
\eea
\end{widetext}
We recall that the fermionic and bosonic Matsubara frequencies are given by
\be
\omega_n = 2\pi T(n+1/2)\;,\;\Omega_m = 2\pi T \,m\,,
\ee
the boson self-energy is defined as 
\be\label{eq:Pi2def}
\Pi(q)=D^{-1}(q)-(q^2+M_D^2 |\Omega|/q)
\ee 
in terms of the quantum boson propagator $D(q)$, and the quantity $A(\omega)$ is related to the wavefunction renormalization $Z(\omega)$ and self-energy $\Sigma(\omega)$ by
\be\label{eq:Adef}
A(\omega_n) \equiv Z(\omega_n) \omega_n= \omega_n+ \Sigma(\omega_n)\,.
\ee
Details of the derivation of these equations, including the symplectic symmetry breaking pattern of $SU(N)$ due to the gap, are presented in Appendix~\ref{ap:SD}. We stress again that we start with a Landau-damped boson in the UV in order to have large $N$ control and avoid the issues found in~\cite{Lee2009}; the boson self-energy (\ref{eq:Pi2def}) will also have a term of the form $|\Omega|/q$, but suppressed by $1/N$.

The set of equations (\ref{eq:TSD}) appears to have a basic problem. At sufficiently large temperature we expect to be in a disordered phase with vanishing gap. Then the static bosonic mode has $\Pi(\Omega_m=0)=0$ because the poles for its momentum 
integral are on the same side of the complex plane, and plugging this into the fermion self-energy equation gives a logarithmic $\int dq/q$ type infrared divergence from the $m=n$ term in the Matsubara sum. This divergence could be resolved if the gap is nonzero. So we seem to arrive to a puzzle: on the one hand we expect the SD equations to have nonsingular solutions, which happens if the gap is nonzero; but on the other hand we expect the gap to vanish at high temperatures.\footnote{Interestingly, this does not seem to be a theorem: Ref.~\cite{Chai:2020zgq} constructed models in fractional dimensions with ordered phases at high temperatures.} Therefore, the interplay between superconductivity and quantum criticality  appears to be conceptually different from the $T=0$ situation.

In this section we will focus on determining the transition temperature $T_c$ for the onset of the superconducting instability. For this, we start in the normal phase with $\Delta=0$ at sufficiently high temperature, and decrease it until the linearized version of the gap equation in (\ref{eq:TSD}) admits a solution. This is an eigenvalue problem that will determine $T_c$. In Sec.~\ref{sc:non-linear} we will analyze the nonlinear equations. We find that at sufficiently low temperatures, a new branch of nonlinear solutions arises, distinct from the linearized solution at that critical temperature. This nonlinear solution provides an alternative mechanism for resolving infrared divergences by developing a superconducting gap. Evaluating the free energy, however, shows that it is not energetically favorable compared to the normal state. So the extension of the linearized solution discussed here to $T<T_c$ gives the dominant minimum.

\subsection{Review of the normal state}\label{subsec:normal}
 
We begin by briefly reviewing the main results of~\cite{Damia:2020yiu} that are required for our analysis of the self-consistent equations (\ref{eq:TSD}).

Resumming the leading order diagrams involving bosonic quartic interactions $\lambda_\phi \phi^4$ gives a self-consistent boson mass to the static zero frequency mode
\be
D(0,q)= \frac{1}{q^2+m^2_b} \label{eq:staticprop}
\ee
with 
\be\label{eq:mb1}
m_b^2 \approx \frac{\lambda_\phi T}{4\pi}\,\,\log\left(4\pi \frac{(2\pi T M_D^2)^{2/3}}{\lambda_\phi T} \right)\,,
\ee
see Appendix~\ref{subsec:mb}.
A similar mass appears for higher Matsubara modes, but is irrelevant there due to their $z_b=3$ scaling, consistently with the $T=0$ irrelevance of $\phi^4$.

Plugging this modified bosonic propagator into the Schwinger-Dyson equation for the fermion self-energy gives a solution that contains two distinct contributions,
\be\label{eq:Sigmathermal}
\Sigma(\omega_n) = \Sigma_T(\omega_n) + \SigmaN(\omega_n)\,.
\ee
The second term $\SigmaN(\omega_n)$ comes from the Matsubara sum with $m \neq n$ in  (\ref{eq:TSD}). This is the ``quantum'' contribution, for which effects of the thermal mass are negligible, and hence it is governed by the scalings of the quantum critical point. It
takes the form
\be
\frac{\SigmaN(\omega_n)}{{\rm sgn}(\omega_n)} \approx  \Lambda^{1/3}(2\pi T)^{2/3}\left(\zeta(\frac13)-\zeta(\frac13,|n+\frac12|+\frac12)\right) \label{eq:NFLself}.
\ee
At low temperatures $T/|\omega_n| \ll 1$, this asymptotes to the quantum critical point result (\ref{eq:SigmaNFL}).

On the other hand, the thermal term $\Sigma_T$ captures loop effects from exchange of static bosons, whose propagator is \eqref{eq:staticprop}. It is determined by the following equation
\be
\Sigma_T = \frac{g^2 T}{2\pi} \frac{\cosh^{-1}\left(\frac{1}{v}\sqrt{\frac{A_n^2 }{ m_b^2}} \right)}{\sqrt{A_n^2 - v^2  m_b^2}}\label{eq:sigmaTeq}
\ee
where we recall that $A_n=\omega_n+\Sigma(\omega_n)$. This equation can be solved numerically, but for our purpose it is sufficient to point out the analytic behaviors in the different regimes~\cite{Damia:2020yiu}:
\be \label{eq:sigmaT}
\Sigma_T(\omega_n) \approx \left\lbrace 
\begin{array}{ccc}
\sqrt{ \log (\frac{g^2}{\lambda_\phi})\frac{g^2 T}{4\pi}} & , & \omega_m < \Lambda_T \\
\frac{g^2 T}{\SigmaN(\omega_n)} & , & \Lambda_T < \omega_n < \Lambda \\
\frac{g^2 T}{\omega_n} & , & \Lambda < \omega_n
\end{array}
\right.
\ee
where $\Lambda_T$ is the frequency scale at which $\SigmaN$ becomes comparable with $\Sigma_T$, namely
\be
\Lambda_T \approx  \left( \frac{1}{4\pi}  \log (\frac{g^2}{\lambda_\phi})\right)^{3/4}\frac{g^{3/2}T^{3/4}}{\Lambda^{1/2}}\,.
\label{eq:LambdaT}
\ee
We note that as $\lambda_\phi/g^2 \to 0$, $\Sigma_T$ diverges logarithmically; this is the original infrared divergence from exchange of massless static bosons in $d=2$.

The existence of the thermal regime for $\omega_n < \Lambda_T$ where $A_n \sim \Sigma_T \sim (g^2 T)^{1/2}$ leads to a new region in the phase diagram of the normal state; this violates the quantum critical scaling, that would instead require $\Sigma \sim T^{2/3}$~\cite{Damia:2020yiu}. The self-energy with thermal contribution no longer vanishes at the first Matsubara frequencies,
\be\label{eq:nonfirst}
\Sigma(\pm \pi T) = \pm \Sigma_T \neq 0\,,
\ee
a fact that will modify the superconducting dynamics considerably. Such contributions decrease with $T$ more slowly than the quantum ones, and hence will dominate at low temperature.

\subsection{Failure of Anderson's cancellation for $N>1$}\label{subsec:anderson}

Let us now analyze the superconducting gap at $T=T_c$. The linearized gap equation reads (See Appendix~\ref{ap:SD})
\ba\label{eq:tDelta1}
\t \Delta(\omega_n) &= \frac{1}{N}  \Sigma_T(\omega_n)\,\frac{\t \Delta(\omega_n)}{A(\omega_n)} \nonumber\\
&+ \frac{\xi}{N} \pi T \sum_{m \neq n}\, \frac{1}{|m-n|^{1/3}}\,\,\frac{\t \Delta(\omega_m)}{A(\omega_m)}\,.
\ea
where we recall that
\be
A(\omega_n) = \omega_n + \Sigma_T(\omega_n) + \Sigma_{NFL}(\omega_n)
\ee
and we have combined the coupling and temperature into a dimensionless constant
\be\label{eq:xidef}
\xi = \frac{g^2}{3 \sqrt{3}\pi v (2\pi T M_D^2)^{1/3}}=\frac{2}{3} \left(\frac{\Lambda}{2\pi T}\right)^{1/3}\,.
\ee
We will now argue that thermal effects from static bosons do not cancel for $N>1$. A similar non-cancellation was observed in $d=3-\epsilon$ in~\cite{Wang:2017teb}.

The gap equation for the physical gap
\be\label{eq:Delphys1}
\Delta(\omega_n) = \frac{\omega_n}{A(\omega_n)}\,\t \Delta(\omega_n)\,.
\ee
reads
\ba
&\left( 1+ \left(1- \frac{1}{N} \right) \frac{\Sigma_T(\omega_n)}{\omega_n}+\frac{\Sigma_{NFL}(\omega_n)}{\omega_n} \right) \Delta(\omega_n)\nonumber\\
& \qquad \qquad = \frac{\xi}{N}\, \pi T\, \sum_{m \neq n}\, \frac{1}{|m-n|^{1/3}}\, \frac{\Delta(\omega_m)}{|\omega_m|}\,.
\ea
We see that thermal and quantum NFL corrections are represented via the quantity
\be
\bar A(\omega_n) = \omega_n +\left(1- \frac{1}{N} \right) \Sigma_T(\omega_n) + \Sigma_{NFL}(\omega_n)\,.
\ee
This shows that thermal effects cancel only for $N=1$.

Defining the analog of $\t \Delta(\omega_n)$ in (\ref{eq:Delphys1}) but in terms of the relevant quantity $\bar A(\omega_n)$,
\be
\Delta(\omega_n) = \frac{\omega_n}{\bar A(\omega_n)}\,\bar \Delta(\omega_n)
\ee
we have the following form of the gap equation:
\be\label{eq:barDelta}
\bar \Delta(\omega_n) = \frac{\xi}{N}\, \pi T\, \sum_{m \neq n}\, \frac{1}{|m-n|^{1/3}}\, \frac{\bar \Delta(\omega_m)}{|\bar A(\omega_m)|}\,.
\ee
This is of the form of the original gap equation dropping the $m =n$ term but, crucially, $\bar A(\omega_m)$ here contains the effects from $\Sigma_T$. This has a very large effect because of the different scaling $\Sigma_T \sim T^{1/2}$ while $\Sigma_{NFL} \sim T^{2/3}$.

The rescaled gap equation (\ref{eq:barDelta}) is superficially similar to that analyzed in recent works on the $\gamma$ model~\cite{2016PhRvL.117o7001W, 2019PhRvB..99n4512W, 2019arXiv191201797C, Abanov_2020, Wu_2020}.
However, the central difference is that $\bar A(\omega_m)$ includes here thermal effects via $\Sigma_T$, which are not taken into account in those works. This is why or results will be quite different.

\subsection{Numerical results for $T_c$}\label{subsec:linear}

It is convenient to perform numerical calculations in terms of dimensionless variables
\be\label{eq:dimensionless}
\begin{array}{ccc}
\hat \omega_n = (\pi T)^{-1}\omega_n & , & \hat\Sigma_n = (\pi T)^{-1}\Sigma(\omega_n) \\
\hat m^2 = (\pi T)^{-2}m^2_b & , & \hat\Delta_n = (\pi T)^{-1}\tilde\Delta(\omega_n)    
\end{array}\,.
\ee
Linearizing the gap equation of (\ref{eq:SD-dim2}) then gives an eigenvalue problem
\be\label{eq:eigenvalues}
v_n= \frac{1}{N} \sum_{m=0}^\infty U_{nm}\,v_m
\ee
for the hermitean kernel
\be\label{eq:U}
U_{nm}= \xi \,\frac{1}{\hat A_n^{1/2}}\, K_{nm}^+\, \frac{1}{\hat A_m^{1/2}}
\ee
with
\be\label{eq:Kp}
K_{nm}^+ = \frac{1- \delta_{nm}}{|m-n|^{1/3}} + \frac{1}{|m+n+1|^{1/3}}\,,
\ee
and
\be
\hat A_n \approx 2n+1 + \hat \Sigma_T+2 \xi \left[\zeta(\frac{1}{3})- \zeta(\frac{1}{3},n+1) \right]\,.
\ee
We note again that the dominant contribution from the thermal NFL effects described in Sec.~\ref{subsec:normal} appears here through $\hat \Sigma_T$ in $\hat A_n$, which in turn affects the kernel $U$ via (\ref{eq:U}). 

A useful way to think about (\ref{eq:eigenvalues})  comes from expanding the free energy to quadratic order in the gap (see Appendix~\ref{ap:free})
\be\label{eq:Fquad}
F_f \sim \sum_{m n} (\hat A_n^{1/2}\,\hat \Delta_n) \, \left( \delta_{nm}-\frac{1}{N} U_{nm}\right)\,(\hat A_m^{1/2}\,\hat \Delta_m)\,.
\ee
Therefore the mass matrix for the gap is 
\be\label{eq:mass-matrix}
M^2= \mathbf 1- N^{-1} U\,.
\ee
As long as $M^2$ has positive eigenvalues, the system is stable under superconductivity.  We will verify that the disordered phase is always stable at sufficiently large $T$. As $T$ is decreased, the superconducting instability will set in for the largest $T$ such that $M^2$ first develops a zero eigenvalue. This defines $T_c$, and (\ref{eq:eigenvalues}) is the equation for a vanishing eigenvalue $M^2 v= 0$. In what follows we will determine $T_c= T_c(\Lambda, g^2, N)$.

The superconducting transition temperature $T_c$ is determined numerically by solving the eigenvalue equation (\ref{eq:eigenvalues}) which, as discussed before, gives a vanishing eigenvalue for the gap mass matrix. However, a faster way to proceed numerically is to look for the largest eigenvalue of $U_{mn}$; this can be done efficiently via variational methods such as Lanczos or Arnoldi algorithms. Let us denote this largest eigenvalue by $\lambda_{max}(T/g^2,\Lambda/g^2)$. Then setting
\be\label{eq:Tceq}
N=\lambda_{max}(T/g^2,\Lambda/g^2)
\ee
gives a vanishing eigenvalue for (\ref{eq:Fquad}), or equivalently it solves (\ref{eq:eigenvalues}). Moreover, since we did this for the largest eigenvalue of $U$, the mass matrix has exactly one zero eigenvalue and the remaining ones are positive. This is the onset of the superconducting instability, and so (\ref{eq:Tceq}) determines $T_c$ given $N$ and $\Lambda/g^2$.

In order to determine the largest eigenvalue of the infinite matrix $U_{mn}$, we set up a numerical routine where the Matsubara indices run up to a cutoff $N_{max}$. For a given choice of $T/g^2$ and $\Lambda/g^2$ we determine the largest eigenvalue for different choices of $N_{max}$, and then extrapolate to $N_{max} \to \infty$. Furthermore, we implement the method of~\cite{Wang:2017teb} that switches the sampling of Matsubara levels from linear to exponential at some given large value of the frequencies. We also check convergence in this choice. This exponential sampling allows us to obtain exponentially small critical temperatures, that is needed in order to characterize the nature of the transition.

\begin{figure}[h]
  \centering
  \includegraphics[width=1.\hsize]{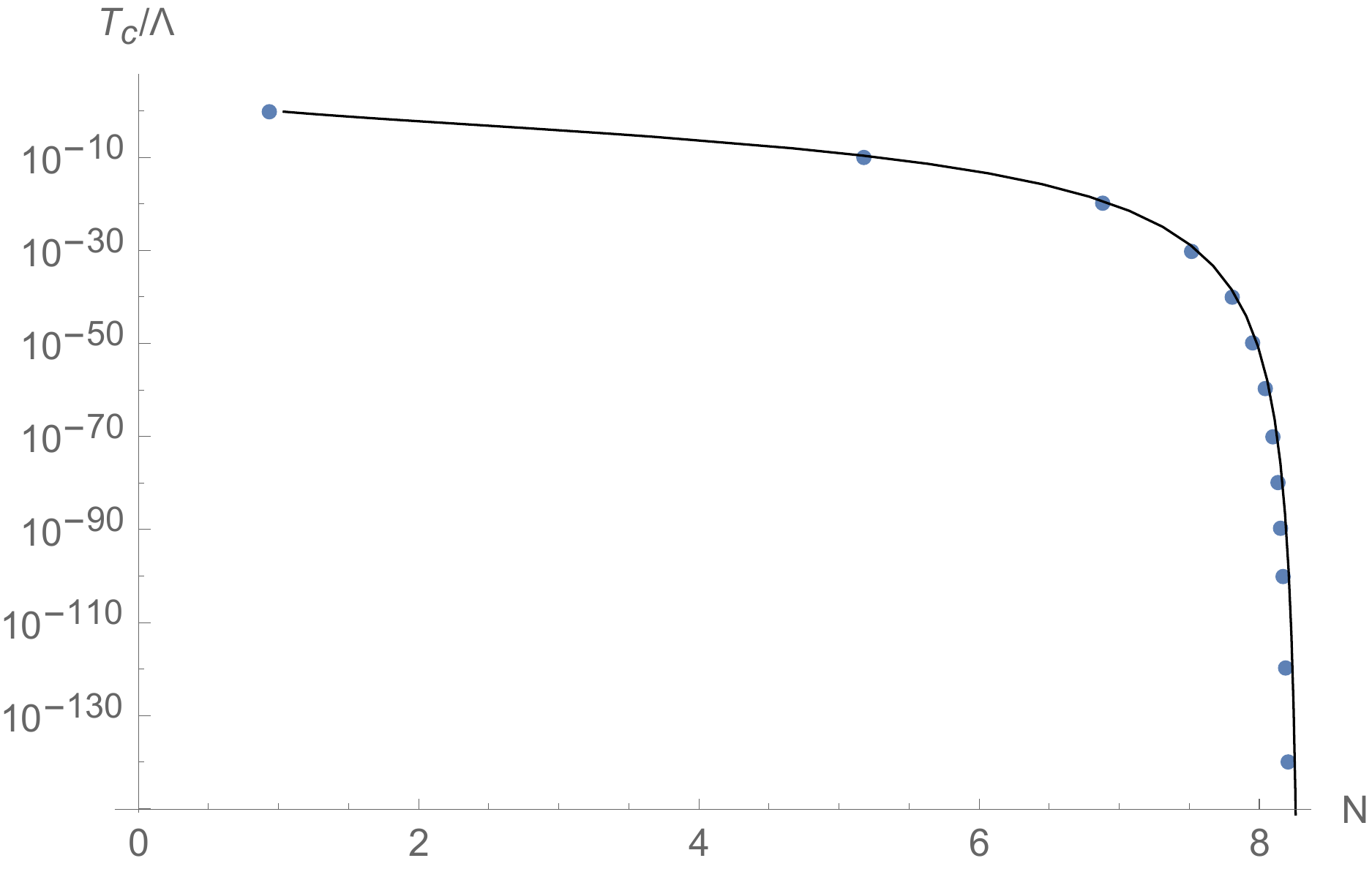}
  \caption{Dimensionless critical temperature ($T_c/\Lambda$) as a function of $N$ for $\Lambda=g^2$. The blue dots have been determined numerically, and agree very well with the semi-analytic prediction of Subsec.~\ref{subsec:semi} (black curve).}
  \label{fig:Tc-N}
\end{figure}

We have verified numerically that $T_c/\Lambda \propto \Lambda/g^2$ up to logarithmic corrections; we explain the origin of this in Sec.~\ref{subsec:semi}. It is then sufficient to set $\Lambda=g^2$. We show our numerical results in Fig.~\ref{fig:Tc-N}. In the next subsection we present a semi-analytic approach that will explain the main features of these results -- the BKT behavior and a parametric difference between $T_c$ and the physical gap.

\subsection{Semi-analytic approach}\label{subsec:semi}

Given the previous numerical results, we need to understand the parametric dependence of $T_c$ on $N$ and $\Lambda/g^2$, as well as the nature of the transition $T_c \to 0$ that is seen around $N \sim 8.3$. With this aim,  let us now develop an approximate approach for solving the gap equation. Since we are interested in small temperatures (in fact, vanishingly small near the putative transition), for our purpose it will be sufficient to work with the continuum approximation to the Matsubara sum,
\be
\tilde\Delta(\omega)= \frac{1}{2N}\int_{\pi T_c}^\infty d\omega' u(\omega,\omega')\frac{\tilde\Delta(\omega')}{A(\omega')}
\label{eq:finiteTSDE}\,.
\ee
The lower integration limit takes into account that the fermionic Matsubara levels start at $\omega=\pi T$; the linearization is valid because we set $T=T_c$. Nonlinear effects will be discussed in Sec.~\ref{sc:non-linear}.
Works with related methods include~\cite{2016PhRvL.117o7001W, Wang:2017teb, Abanov_2020, Wu_2020}.

The first key point is that $T_c$ appears to vanish for $N \gtrsim 8.3$. This is consistent with the $T=0$ analysis of the superconducting instability of Sec.~\ref{sec:model}, were it not for the fact that both the RG and differential equation approaches give a slightly smaller value $N_{cr}=8$. This difference arises because the full eigenvalue problem (\ref{eq:eigenvalues}) -- (\ref{eq:Kp}) that determines $T_c$ includes nonlocal effects that are not taken into account in the local approximation. These arise from terms $m \neq n$ in the Matsubara sum (\ref{eq:eigenvalues}) or, equivalently, from regions in the integral (\ref{eq:finiteTSDE}) with $\omega'$ different from the external frequency.  The difference between the local and nonlocal results is just of a few percent, but a more precise analytic determination of $N_{cr}$ is required in order to understand the transition between the normal and superconducting states.

One way to obtain $N_{cr}$ analytically is to approach $N \to N_{cr}$ from above, in which case $T_c \to 0$, and focus on the range $\omega<\Lambda$. Since the fermion dynamics here is dominated by the quantum critical behavior, we expect a power-law dependence for $\t \Delta(\omega)$. Neglecting subleading corrections from the Fermi-liquid range of the integral in (\ref{eq:finiteTSDE}), the gap equation becomes
\be\label{eq:SDEapprox}
\t \Delta(\omega) \approx \frac{2}{3N}\,\int_{ 0}^\Lambda d\omega' \left(\frac{1}{|\omega'-\omega|^{1/3}}+ \frac{1}{(\omega+ \omega')^{1/3}} \right)\frac{\t \Delta(\omega')}{ (\omega')^{2/3}}
\ee
From (\ref{eq:app-solT0}), we expect a power-law dependence $\t \Delta(\omega) \sim \omega^{-1/6}$ near the transition, so let us parametrize
\be\label{eq:scalingansatz}
\t \Delta(\omega) \approx \frac{1}{\omega^{\frac{1}{6}+ \frac{\nu}{3}}}\;,\; \nu \ll 1\,.
\ee
Performing the integral in the right hand side of (\ref{eq:SDEapprox}) with (\ref{eq:scalingansatz}) and small $\nu$, and matching with the left hand side, gives
\be
N \approx 8.307+32.01 \,\nu^2\,.
\ee
Hence
\be\label{eq:nusmall}
\nu  \approx \pm 0.51 \sqrt{\frac{N}{8.307}-1}\,.
\ee
The square root behavior is the smoking gun signature of the BKT transition: we find two real solutions --corresponding to the UV and IR fixed points (\ref{eq:lambdapm}) of the BCS coupling-- which disappear and go off to the complex plane for $N< 8.307$.
We see then that the critical value from the local approximation is modified to 
\be\label{eq:Ncrnew}
N_{cr} \approx 8.307
\ee
due to nonlocal effects. Recently, Ref.~\cite{2019arXiv191201797C} obtained an analytic expression $N(\nu)$ for all $\nu$; it agrees with (\ref{eq:Ncrnew}) at small $\nu$. For our purpose, (\ref{eq:nusmall}) will be sufficient.

Our approach now will be to derive a differential equation approximation to (\ref{eq:finiteTSDE}), improved by the result (\ref{eq:Ncrnew}) in a way that we explain shortly. We will find that this provides a very good fit for the numerical results, as well as a physical understanding for the behavior of $T_c$.

The local approximation to (\ref{eq:finiteTSDE}) gives again \eqref{eq:zeroTdiff}, but finite $T=T_c$ introduces two modifications: $A(\omega)$ now includes thermal effects from $\Sigma_T$ in (\ref{eq:Sigmathermal}), and the infrared boundary condition becomes
\be
\tilde\Delta'(\pi T_c) = 0\,.
\label{eq:criticalbc}
\ee
This will fix $T_c$. Recall that at $T=0$ we had $\tilde\Delta'(\omega=\Delta_0)=0$ at the physical gap $\Delta_0$. In Fermi-liquid superconductivity, the physical gap and critical temperature are in fact the same up to pre-exponential factors. But will find that, due to NFL thermal effects encoded in $\Sigma_T$, there is a parametric suppression of $T_c$ compared to the gap. So it is important to distinguish both quantities.

As for the $T=0$ case, it is useful to use the dimensionless variable $x$ given by \eqref{eq:x},
for which we have
\be
A(\omega)= \Lambda e^{-2x}(e^{-x}+1+e^{2(x-x_T)})\,.
\ee
The quantity $x_T$ is defined by 
\be\label{eq:xTdef}
\Lambda_T=\Lambda e^{-3x_T}\,,
\ee 
and $\Lambda_T$ was given in \eqref{eq:LambdaT}.

The differential equation together with the boundary conditions are
\be
\tilde\Delta'(x)-\tilde\Delta''(x)=\frac{\t g_1}{e^{-x}+1+e^{2(x-x_T)}}\tilde\Delta(x)
\label{eq:x-diff2}
\ee
\be\label{eq:diff2bc}
\tilde\Delta'(x_c)=0 \,\, , \,\, \tilde\Delta(x\to-\infty)\sim e^{x} 
\ee
where $x_c$ is just $T_c$ in terms of the $x$-variable, $\pi T_c=\Lambda e^{-3x_c}$ and the second equation above corresponds to the boundary condition in the UV. 

The strictly local approximation gives $\t g_1=2/N$ as in (\ref{eq:x}), (\ref{eq:x-diffT0}); this would imply $N_{cr}=8$. Instead, here we have kept $\t g_1$ more general so that we can incorporate nonlocal effects. In order to reproduce (\ref{eq:Ncrnew}), we have to choose
\be
\t g_1= \frac{1}{4}\frac{N_{cr}}{N} \approx \frac{1}{4} \frac{8.3}{N}\,.
\label{eq:g1-Ncr}
\ee
Although we won't need further details, we remark that, following an approach similar to~\cite{Abanov_2020}, such a modification can be incorporated by keeping a region $\omega'$ around $\omega$ in the integral where $u(\omega, \omega')$ is not simplified by the local approximation. Rather, the integral is evaluated by saddle point in this region.

It is now easy to solve (\ref{eq:x-diff2}) and (\ref{eq:diff2bc}) numerically, given (\ref{eq:g1-Ncr}). The resulting $\t \Delta(\omega)$ is shown as the black curve in Fig.~\ref{fig:Tc-N}, exhibiting an excellent agreement with the numerical finite temperature data.

It remains to determine the parametric dependence of $T_c$ on $\Lambda/g^2$ and $N$. For this purpose, it is sufficient to construct approximate solutions in appropriate regimes of the frequency, subsequently imposing continuity and differentiability at the gluing points. There are three distinct regions: the high frequency regime $-\infty <x<0$, dominated by the first term in the denominator of \eqref{eq:x-diff2}; the intermediate regime $0<x<x_T$, determined by the quantum NFL behavior (second term in the denominator of (\ref{eq:x-diff2})); and the low frequency region $x_T<x$, dominated by thermal effects from $\Sigma_T$ (third term in that denominator). The main effect of this last contribution is to suppress exponentially the right hand side of \eqref{eq:x-diff2}, making the solution approach a constant for $ x> x_T$. 

All in all, the piecewise solution now reads
\be
\t \Delta(x) \approx \left\lbrace 
\begin{array}{lcc}
e^{\frac x 2}J_1(2\sqrt{ \t g_1}e^{\frac x 2}) & , & -\infty < x < 0 \\
C_1 e^{\frac x 2} \cos\left(\sqrt{\t g_1-\frac{1}{4}}x+\phi \right) & , & 0<x<x_T \\
D_1 & , & x>x_T 
\end{array}
\right.
\label{eq:app-sol}
\ee
For high frequencies we have retained only the decaying solution $\Delta\sim\omega^{-1/3}$ as $\omega\to\infty$, as imposed by the UV boundary condition. The overall scale here is not fixed, since we are looking at the linearized problem. On the other hand, the constants $C_1$, $\phi$ and $D_1$ are fixed by imposing continuity of $\Delta(x)$ and $\Delta'(x)$ at the matching points, thus becoming functions of $\t g_1$.

The main new effect is that the solution stabilizes to a constant value for $x> x_T$ or, equivalently, for $\omega < \Lambda_T$. This happens for frequencies in the range such that the self-energy is completely dominated by the thermal piece $\Sigma_T$. As a result, thermal effects shift the IR boundary condition \eqref{eq:criticalbc} from $x_c$ to $x_{T_c}<x_c$, namely
\be
\t \Delta'(\Lambda_{T_c})=0\,.
\ee
This behavior was observed before in the $d=3-\epsilon$ analysis of~\cite{Wang:2017teb}; the physical consequences will be here greatly amplified compared to~\cite{Wang:2017teb} because quantum and thermal effects are much stronger in $d=2$ dimensions.

The physical gap $\Delta_0$ is the frequency at which $\t \Delta(\omega)$ stabilizes to a constant, and hence here
\be\label{eq:Delta0final}
\Delta_0 = \Lambda_{T_c}\,.
\ee
This scale sets the onset of superconductivity.
The last step is to determine $T_c$ in terms of the parameters of the theory. For this, it is easiest to recognize that we have already solved an equivalent problem in (\ref{eq:x-diffT0}) and  (\ref{eq:app-solT0}) -- we only need to replace $g_1$ there by the improved $\t g_1$ in (\ref{eq:g1-Ncr}), as well as identify $x_0$ of that problem with $x_{T_c}$ in our current analysis. The result is
\be
x_{T_c}\approx \frac{2\pi}{\sqrt{4 \t g_1-1}}\,.
\ee
Knowing $x_{T_c}$ determines $\Lambda_{T_c}=  \Lambda e^{-3 x_{T_c}}$. Recalling its expression (\ref{eq:LambdaT}) in terms of the temperature then gives $T_c$, and furthermore plugging into (\ref{eq:Delta0final}) fixes the gap. The final results are
\bea\label{eq:Tc-g1}
\frac{T_c}{\Lambda} & \approx & \frac{4\pi}{\log \frac{g^2}{\lambda_\phi}}\,\frac{\Lambda}{g^2}\,\exp\left(- \frac{8\pi}{\sqrt{N_{cr}/N-1}}\right) \nonumber\\
\frac{\Delta_0}{\Lambda} & \approx &  \exp\left(- \frac{6\pi}{\sqrt{N_{cr}/N-1}}\right)\,.
\eea

Let us highlight four consequences of these results.
\begin{itemize}
\item This reveals an infinite order BKT-type transition as $N \to N_{cr}$; conceptually, this is in agreement with previous  RG and Eliashberg studies  in $d=3-\epsilon$ at $T=0$~\cite{Raghu:2015sna, Wang:2016hir} and at finite $T$~\cite{Wang:2017teb}. This analytic behavior is in excellent agreement with the numerical results, as illustrated above in Fig.~\ref{fig:Tc-N}. 
\item Both $T_c$ and $\Delta_0$ depend exponentially on a function of $N$. While the $N$-dependence shown in (\ref{eq:Tc-g1}) is corrected away from $N \sim N_{cr}$, the numerical results show large suppressions even for $N \sim 1$. 
This should be contrasted with the naive expectation that in $d=2$ the values of $(T_c, \Delta_0)$ should be determined by $\Lambda$ and/or $g^2$ with some power-law dependence but without exponentials. Instead, we have obtained a large suppression that comes from strong quantum NFL effects.
\item Thermal fluctuations introduce a parametric scale separation $T_c \ll \Delta_0$ that originated from the premature stabilization of the gap at $\Lambda_{T_c}$. $T_c$ is \textit{exponentially smaller} than $\Delta_0$, except at the smallest $N \sim 1$. On top of this, there is an extra suppression in $T_c$ by $(\Lambda/g^2)$ compared to $\Delta_0$. The hierarchy $T_c/\Lambda_0 \ll 1$ provides a distinct signature of non-Fermi liquid superconductivity. 
\item Finally, in the limit $\lambda_\phi/g^2 \to 0$, we find $T_c /\Lambda \to 0$ from the inverse logarithm in (\ref{eq:Tc-g1}). This shows explicitly how large thermal corrections from virtual static bosons tend to destroy superconductivity.\footnote{We thank A. Abanov and A. Chubukov for discussions on this point.}
\end{itemize}

Strictly at $N=1$ thermal effects cancel, as shown in Sec.~\ref{subsec:anderson}. These results apply for $N>1$. They illustrate very explicitly the general points of Sec.~\ref{sec:thermal} on the conceptual distinction between $N=1$ and $N>1$ in NFLs.

\subsection{Summary and phase diagram}

\begin{figure}[h]
\centering
\def\svgwidth{8cm}
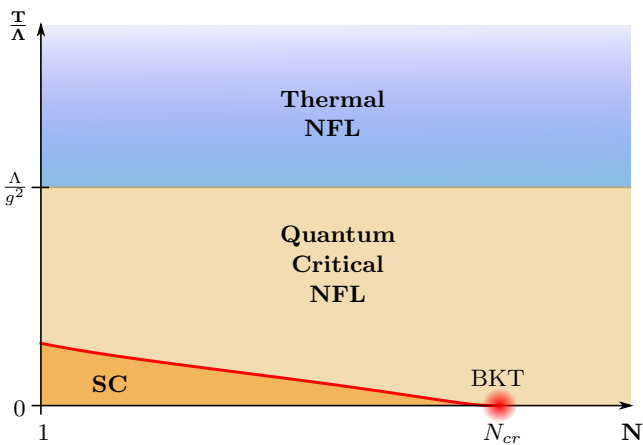
\caption{Schematic phase diagram, the lower half being the main focus of this work, {\it i.e.} temperatures below $\Lambda^2/g^2$. The red line represents the second-order transition to the superconducting state, with critical temperature well-approximated by (\ref{eq:Tc-g1}). The point $(T=0,N=N_{cr})$ where the transition line ends displays BKT criticality. Beyond that point, the system flows to a stable NFL with critical (finite) couplings.}
\label{fig:phase-diag}
\end{figure}

This ends our analysis of superconductivity around $T = T_c$, so it is useful to summarize the picture that emerges for the NFL dynamics. A schematic phase diagram is presented in Fig.~\ref{fig:phase-diag}.

\vskip 5mm

For $N>N_{cr}$ the system does not superconduct, but rather displays a stable \textit{naked quantum critical point} at $T=0$. The BCS coupling also reaches a stable fixed point.
The finite $T$ dynamics of this normal state has been recently discussed in~\cite{Damia:2020yiu}. The traditional picture of a quantum critical region determined by $T=0$ scaling laws is strongly modified by thermal effects that become important at $T > \Lambda^2/g^2$ (the blue region in Fig.~\ref{fig:phase-diag}). In this range, the fermion Green's function scales like $G_F \sim T^{-1/2}$ instead of $G_F \sim T^{-2/3}$. 

At $N= N_{cr}$, numerical results are consistent with an infinite order transition, with BKT scaling (\ref{eq:Tc-g1}). Physically, the RG provides a compelling explanation: the IR stable QCP annihilates against an unstable fixed point. This is represented by the $\beta$-function (\ref{eq:BCSRG}).

For $N< N_{cr}$, the ground state is a superconductor. The line $T=T_c(N)$ of second order phase transitions ends at the BKT transition at $N=N_{cr}$. We have found that this NFL superconductor is quite different from BCS expectations. There is an important suppression both of $T_c$ and the physical gap $\Delta_0$ even for $N \sim 1$. Furthermore, a clear phenomenological signature is given by the parametric suppression $T_c \ll \Delta_0$ observed numerically and explained analytically in (\ref{eq:Tc-g1}). 

It is quite plausible that the large $N$ results are not quantitatively correct down to $N \sim 1$. But we expect the previous qualitative picture to survive in this case. In fact, at small $N$ we expect stronger quantum corrections and hence further enhancement of NFL effects, but it would be interesting to analyze this further. This key distinction between $N=1$ and $N>1$ motivates future numerical analyses of this or related NFL models.

\section{Non-linear superconductivity and infrared divergences}\label{sc:non-linear}

So far we have studied the competition between NFL effects and superconductivity at finite temperature, by focusing on $T=T_c$ and linearizing the gap equation. In this last section we will analyze the full Schwinger-Dyson equations, including the nonlinear gap equation. The usual type of nonlinear solution obtains from extending the linearized solution of the previous section to $T< T_c$. The dynamics here is that expected from a gapped system with spontaneous symmetry breaking. Instead, our goal is to determine whether other solutions might exist at the nonlinear level. 

There are two motivations to carry out this analysis. The first is that the linearized solution used a self-consistent thermal boson mass to cure the infrared divergences. However, one can envision that the superconducting gap itself could give rise to a bosonic mass and hence eliminate the thermal divergences. The dependence on the gap would then be necessarily non-analytic and requires solving the nonlinear Schwinger-Dyson equations. The other motivation comes from ``first Matsubara law'' solutions found in the closely related $\gamma$ model~\cite{2016PhRvL.117o7001W, 2019PhRvB..99n4512W, 2019arXiv191201797C, Abanov_2020, Wu_2020}. These solutions are driven predominantly by superconducting gaps along the first few Matsubara modes, which is quite different from the linearized solution above. Our theory differs from that in~\cite{2016PhRvL.117o7001W, 2019PhRvB..99n4512W, 2019arXiv191201797C, Abanov_2020, Wu_2020} by inclusion of thermal effects, but it is nevertheless natural to ask if similar solutions exist here.

This analysis will be more involved, so let us here summarize our findings and then describe the strategy we will follow. First, at sufficiently large $T$ (we make this precise below), we find that a gap does not develop at nonlinear level. In this case the mechanism for resolving IR divergences is still via a self-consistent boson mass as in~\cite{Damia:2020yiu}. This is fortunately consistent with the standard intuition that at high temperatures the normal state should be stable. On the other hand, at small temperatures we do find that a consistent solution with a nonlinear gap develops, and that this gives rise to a bosonic mass that is generically larger than the self-consistent contribution (\ref{eq:mb1}). This provides then an interesting mechanism for resolution of thermal divergences. To determine the fate of this new branch of solutions, we analyze the Luttinger-Ward free energy, finding that this solution in fact has higher free energy than the normal state. As a result, we conclude that this solution does not appear to be relevant for the finite $T$ dynamics of the theory. Towards the end of the section we will comment on the connection with the nonlinear solution of~\cite{2016PhRvL.117o7001W, 2019PhRvB..99n4512W, 2019arXiv191201797C, Abanov_2020, Wu_2020}.

\subsection{Strategy}

As discussed above, extending the linearized solution to $T< T_c$ will give a gapped superconducting state where $\Delta(\omega)$ stabilizes to $\Delta_0$ as $\omega \to 0$. The linearized solution relied on the thermal mass (\ref{eq:mb1}) to make static boson exchanges finite. On the other hand,
the bosonic mass $m^2$ also receives contributions from the gap, which at one loop read
\be\label{eq:mDelta}
m^2_\Delta = \frac{g^2 k_F T}{2 v N}\sum_{n=0}^\infty\frac{\tilde\Delta_n^2}{(A_n^2+\tilde\Delta^2_n)^{3/2}}\,,
\ee
as derived in (\ref{eq:gap-mass}). It is then in principle possible to have a different branch of solutions at the nonlinear level
if this mass contribution can become larger than (\ref{eq:mb1}). So we will assume $m_\Delta^2 \gg m_b^2$ and look for nontrivial nonlinear solutions.

From the dynamical exponent $z_f =3/2$ of the $T=0$ theory,
we expect that in this new solution 
\be\label{eq:expect}
\Sigma(\omega_n) \sim \Lambda^{1/3} T^{2/3}\;,\;\t \Delta(\omega_n) \sim \Lambda^{1/3} T^{2/3}\,,
\ee 
for $T < \Lambda$,
up to small  corrections suppressed by $1/\Lambda$. This should be contrasted with the results in previous sections, where the violation of the $T=0$ scalings was allowed because of the dangerously irrelevant coupling $\lambda_\phi$. In this section, we are assuming that effects from $\lambda_\phi$ are subleading, since we are interested in new nonlinear solutions.

Having understood what type of temperature scaling we are after, it remains to obtain the dependence on the Matsubara level $n$. It is convenient to switch to dimensionless variables (\ref{eq:dimensionless}).
We look for solutions of the nonlinear Schwinger-Dyson equations,
\bea\label{eq:SD-dim3}
\hat A_n &=&\hat \omega_n + \hat f_n \frac{\hat A_n}{\sqrt{\hat A_n^2 + \hat \Delta_n^2}}+ \xi \sum_{m=0}^\infty K^-_{nm} \frac{\hat A_m}{\sqrt{\hat A_m^2 + \hat \Delta_m^2}} \nonumber\\
\hat \Delta_n &=&\frac{1}{N} \hat f_n \frac{\hat \Delta_n }{\sqrt{\hat A_n^2 + \hat \Delta_n^2}}+\frac{1}{N}\xi \sum_{m=0}^\infty K^\pm_{nm} \frac{\hat \Delta_m }{\sqrt{\hat A_m^2 + \hat \Delta_m^2}} \nonumber \\
\eea
derived in Appendix~\ref{ap:SD}, see \eqref{eq:SD-dim2}. The function $\hat f_n$ is a generalization of $\Sigma_T$ in (\ref{eq:sigmaTeq}) that includes the SC gap,
\be\label{eq:fn2}
\hat f_n \equiv \frac{g^2 }{2\pi^3 T} \frac{\cosh^{-1}\left(\frac{1}{v}\sqrt{\frac{\hat A_n^2+ \hat \Delta_n^2}{\hat m^2}} \right)}{\sqrt{\hat A_n^2+ \hat \Delta_n^2 - v^2 \hat m^2}}\,,
\ee
see  \eqref{eq:fn}. 

Therefore we need to solve the coupled equations (\ref{eq:SD-dim3}), which include nonlinear terms both from the square roots there, as well as from the infrared contributions encoded in $\hat f_n$. This is in general quite hard, because we are dealing with an infinite number of coupled nonlinear equations. We will look for these solutions numerically by truncating to a finite number of Matsubara levels. But before doing this, we will show that the full solution is dominated by the first Matsubara modes. This is similar to the first Matsubara law solution of~\cite{2016PhRvL.117o7001W, 2019PhRvB..99n4512W, 2019arXiv191201797C, Abanov_2020, Wu_2020}, but there are differences and we will compare with those works in Sec.~\ref{subsec:firstM}.

So we obtain the solution explicitly keeping only the first two modes, and then extend it to include additional Matsubara frequencies. While this is approximate (and will be later on improved with numerical results), it allows us to extract the main physical properties of this nonlinear regime. 

\subsection{Approximate truncated solution}\label{sc:gaptrunc} 

The Schwinger-Dyson equations restricted to $n=0$ (i.e. the first Matsubara frequencies $\pm \pi T$) read
\bea\label{eq:A0Delta0eq}
\hat A_0&=& \hat \omega_0 + (\hat f_0 - \xi) \,\frac{\hat A_0}{\sqrt{\hat A_0^2+ \hat \Delta_0^2}} +\xi \nonumber\\
\hat \Delta_0&=&\frac{1}{N}  (\hat f_0 + \xi) \,\frac{\hat \Delta_0}{\sqrt{\hat A_0^2+ \hat \Delta_0^2}} \,.
\eea
Here $\hat f_0$ is given by (\ref{eq:fn2}) with mass restricted to the first mode,
\be\label{eq:app-gapmass}
\hat m^2 \approx \frac{g^2 k_F }{2 \pi^3 v N T^2 }\frac{\hat\Delta_0^2}{(\hat A_0^2+\hat\Delta^2_0)^{3/2}}\,.
\ee
We also note the last term $\xi$ in the equation for $\hat A_0$ -- it is the NFL self-energy part that comes from $\sum_{m \ge 1} K^-_{mn}$ in (\ref{eq:SD-dim3}) after setting $\hat \Delta_{m \ge 1}=0$ there.

At large $N$, we find an approximate solution by neglecting $\hat f_0$,
\be \label{eq:nlsolution}
\hat A_0 \approx \frac{1+\xi}{1+N} \,\, , \,\, \hat\Delta_0 \approx \sqrt{\frac{\xi^2}{N^2}-\left(\frac{1+\xi}{1+N}\right)^2}\,.
\ee
This exists as long as the discriminant is positive, that is
\be
\xi > N \label{eq:exist}\,.
\ee
From the definition of $\xi$ given in (\ref{eq:xidef}), this defines a temperature scale $T_{NL}$
\be\label{eq:TNL}
T_{NL} \approx \frac{4}{27\pi}\, \frac{\Lambda}{N^3}
\ee
so that the nonlinear solution exists for any $N$ at sufficiently low temperatures
\be\label{eq:Tsmall}
T < T_{NL}\,.
\ee
In this range, the solution approximates to
\be
\hat A_0 \approx \frac{\xi}{N}\;,\;\hat \Delta_0^2 \approx 2 \frac{\xi^2}{N^3} \,, \label{eq:truncsol}
\ee
or, translating to the original dimensionful quantities, 
\be
A_0 \approx \frac{1}{3N} \Lambda^{1/3} (2\pi T)^{2/3}\;,\; \t \Delta_0 = \frac{\sqrt 2}{3 N^{3/2}} \Lambda^{1/3} (2\pi T)^{2/3}\,.
\ee
This agrees with the expected scaling (\ref{eq:expect}) of the $T=0$ theory.

We see that the gap is subdominant compared to $\hat A_0$, and one can verify that the initial hypothesis $\hat f_0 \ll \xi$ is consistently satisfied. We will discuss shortly what happens as $\xi \to N$. In obtaining this solution, we have neglected the thermal boson mass and contributions from higher Matsubara modes. We now argue that this is justified.

From (\ref{eq:app-gapmass}) and (\ref{eq:mb1}), 
the ratio between the gap-induced and thermal boson masses is
\be
\frac{m^2_\Delta}{m^2_b}\approx \frac{g^2 k_F}{v \lambda_\phi  N}\frac{1}{\Lambda^{1/3}T^{2/3}}  \approx \frac{g^2 k_F}{v \lambda_\phi \Lambda N}\xi^2\,.
\label{eq:mratio}
\ee
Since $\xi > N$ here, the thermal mass can be neglected as long as
\be\label{eq:lambdaphi-cond}
\frac{\lambda_\phi}{g^2} < N \frac{k_F}{v \Lambda}\,.
\ee
This is easily satisfied because $N > 1$ and $k_F \gg \Lambda$.

Regarding contributions from the higher Matsubara modes $\hat \Delta_{n}$, we note that they cannot be strictly zero because $\hat \Delta_0$ sources them. Indeed, keeping only $\hat \Delta_0$ in the right hand side of (\ref{eq:SD-dim3}) gives a nonvanishing result
\be
\hat \Delta_n \approx \frac{\xi}{N}{K}^+_{n0} \frac{\hat\Delta_0}{\sqrt{\hat A_0^2+\hat\Delta_0^2}} \label{eq:highfreq}\,.
\ee
In order to evaluate the backreaction of $\hat \Delta_n$ on $\hat \Delta_0$, we need $\hat A_n$. Since already $\hat \Delta_0$ was much smaller than $\hat A_0$, we can neglect the gap altogether in the Schwinger-Dyson equation for the fermion self-energy, obtaining
\be
\hat A_{n \ge 1} =\hat \omega_n+\xi \sum_{m=0}^\infty K^-_{nm} = \hat \omega_n + \hat\Sigma_{\rm NFL}^n
\label{eq:tAn}
\ee
where $\hat\Sigma_{\rm NFL}^n=(\pi T)^{-1}\SigmaN(\omega_n)$ with $\SigmaN(\omega_n)$ given in \eqref{eq:NFLself}.  As happened above with $\hat f_0$, effects from $\hat f_n$ are also negligible, and that is why we have not included such thermal self-energy terms here. Using  (\ref{eq:highfreq}) and (\ref{eq:tAn}), the terms with $m \ge 1$ give a relative contribution to the $n=0$ gap equation
\be
\frac{\delta \hat\Delta_0}{\tilde \Delta_0}\approx \frac{\xi}{N}\sum_{m\geq 1}\frac{(K^+_{0m})^2}{\hat A_m}\,.
\ee
Since $\hat A_m\sim \xi m^{2/3}$ for $m\geq 1$, and the sum is convergent, this is therefore suppressed in the large $N$ limit, as promised. 

\begin{figure}[h!]
\begin{center}  
\includegraphics[width=.4\textwidth]{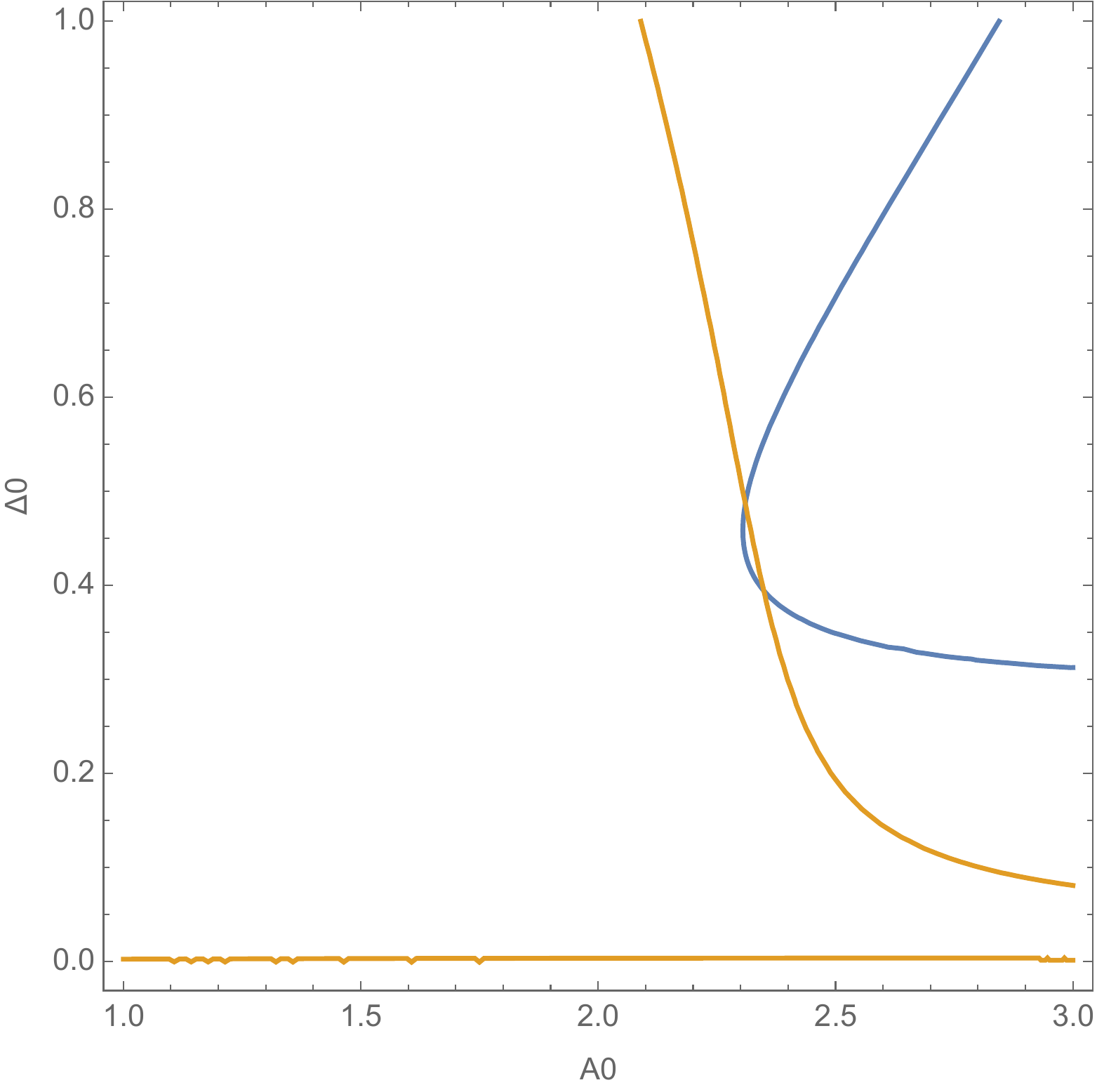}
\caption{Contour plots eq$=0$ for the $A$ equation (blue) and gap equation (orange) for $T=3.1 \times 10^{-9}, N=10, g=M_D=1$, with the two solutions annihilating.This happens at a nonzero value of $\hat \Delta_0$. The constant $\hat \Delta_0=0$ line at the bottom intersects with the blue contour (the $\hat A$ equation) for much larger values of $\hat A_0$, not shown here.}
\label{fig:contour1}
\end{center}  
\end{figure}

It remains to determine what happens as $\xi \to N$. In this limit, $\hat f_0$ can no longer be neglected, because it diverges when $\hat \Delta_0 \to 0$ as predicted by (\ref{eq:nlsolution}). So $\hat f_0 \ll \xi$ will break down somewhere before $\xi =N$. Therefore, thermal effects encoded in $\hat f_0$ (and the $\hat f_n$ more generally) begin to play an important role. The system of equations (\ref{eq:A0Delta0eq}) does not admit a simple analytic solution in this case. But it is not hard to solve it numerically. The basic outcome is that the solution still ceases to exist at a value  of order $\xi \approx N$, but the main difference is that this happens at a finite $\hat \Delta_0$, and not at $\hat \Delta_0=0$ as in (\ref{eq:nlsolution}). This can be seen from the contour plots of the two equations (\ref{eq:A0Delta0eq}) in the $(\hat A_0, \hat \Delta_0)$ plane. An example is given in Fig.~\ref{fig:contour1}. This reveals three solutions. The first one has $\hat \Delta_0=0$ and $\hat A_n$ is the usual normal state result; in the figure, $\hat \Delta_0=0$ appears as the constant orange line at the bottom, which intersects the contour line of the $\hat A$ equation to the far right. This is not shown here because we focus on the range with nonlinear solutions in the plot. Besides the normal state solution, this reveals two other nonlinear solutions, from the two contour intersections in the figure. As $T \to T_{NL}$ the two solutions annihilate, at this order. This happens at finite $\hat \Delta_0$ as seen in the plot. It would be interesting to understand in more detail this mechanism of annihilation, but this is outside the scope of this work (especially given that the solution will not be thermodynamically favored).

To summarize, we have found a nonlinear solution at $T< T_{NL}$, where the gap has the profile (\ref{eq:nlsolution}) and (\ref{eq:highfreq}). In this solution, the thermal boson mass is subdominant due to (\ref{eq:lambdaphi-cond}), and the formation of a gap provides a nonperturbative resolution of thermal divergences.  This solution exists even at $N>N_{cr}$, for which the linearized analysis predicts no superconducting instability. The solution disappears discontinuously for $T \to T_{NL}$. It remains to determine if it is energetically favorable compared to the normal state. We will analyze this point in Subsec.~\ref{sc:inter}. But before turning to this, let us present numerical results for the complete Schwinger-Dyson equations that verify our current approach.

\subsection{Numerics}\label{sc:num-check}

The main conclusions from the previous analysis, namely the scaling (\ref{eq:expect}) and the disappearance of the nonlinear solution as $T \to T_{NL}$, have been verified numerically. We have done this using numerical methods for finding roots of nonlinear systems. We solve the non-linear system \eqref{eq:SD-dim3} up to a given maximum frequency number $n_{max}$ and check convergence by varying this parameter. The truncated solution is used as an input for local searches of solutions in the full system.

\begin{figure}[h!]
\begin{center}  
\includegraphics[width=.5\textwidth]{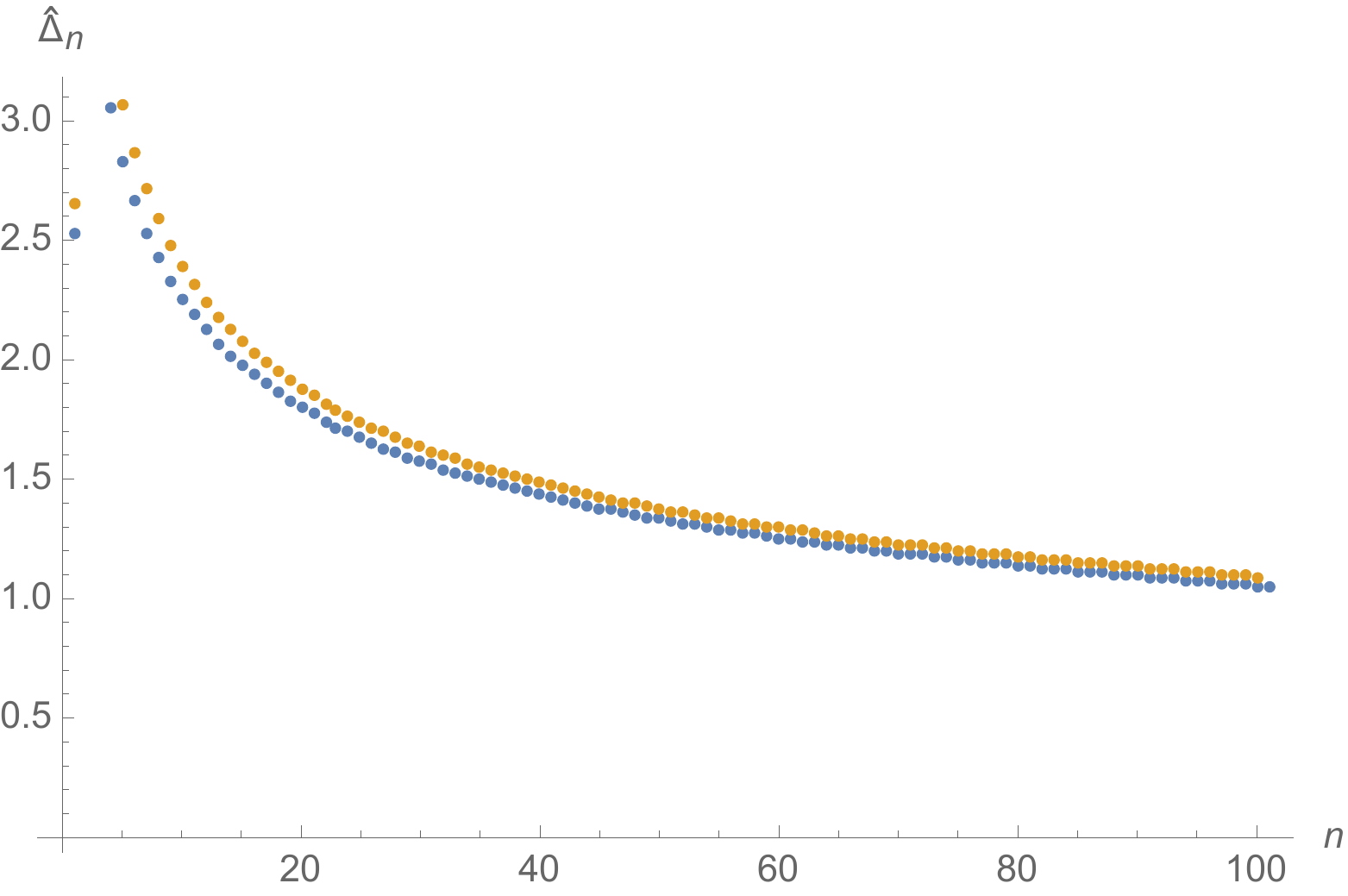}
\caption{Numerical solution $\hat \Delta_n$ to  \eqref{eq:SD-dim3} with $n_{max}=100$ (dots in blue), and truncated solution (orange). The parameters are $N=20,\,\Lambda=g^2=1, \lambda_\phi=1/10, T=10^{-8}$.}
\label{fig:FM}
\end{center}  
\end{figure}

An example is presented in Fig.~\ref{fig:FM}. We chose $N=20$, so we expect the truncated result to give a good approximation, and indeed, this is what we find. Decreasing $N$ but still keeping $T$ low gives solutions that start to deviate from the truncated approximation, but the basic $T^{2/3}$ scaling of $\t \Delta_n$ is respected, as expected.

Increasing the temperature, the numerical solution ceases to exist for $T \sim T_{NL}$ as in (\ref{eq:TNL}). This is consistent with the results from the truncation in the previous section. It should be stressed that we have used local search methods; global methods are much more demanding and would require a different level of numerical analysis, such as that provided by neural networks. So, while at this stage we cannot forbid the existence of other nonlinear solutions, all our results converge to suggest that this is unlikely. Developing efficient methods for solving nonlinear Schwinger-Dyson equations would be very interesting.

\subsection{Thermodynamic fate of the solution}\label{sc:inter}

We have found that accounting for non-linear effects leads to an alternative self-consistent solution to the Schwinger-Dyson equations. This solution is quite different from the normal state at $T>T_c$; it involves a mass term induced by fermionic loops in the presence of a self-consistent gap profile. The physical significance of this solution is not completely clear, particularly regarding the nature of global symmetry breaking. There are some proposals that similar first-Matsubara law solutions could describe a pseudogap state \cite{Abanov_2020}.

In the regime $T_c< T< T_{NL}$ where the normal state and nonlinear gap solutions exist, one of them will be energetically preferred. We will now address this, by evaluating their free energies with the Luttinger-Ward functional \cite{LuttingerWard:1960, Benlagra:2011, ChubukovHaslinger:2003}. 

Let us focus on the  free energy contribution from the Fermi surface (see  Appendix \ref{ap:free})
\be
F_f = -\frac{NT k_F}{v}\sum_{n=0}^{\infty}\left( \sqrt{A_n^2+\t \Delta_n^2}+\frac{|\omega_n| |A_n|}{\sqrt{A^2_n+\t \Delta_n^2}} \right)\label{eq:freeferm}\,.
\ee
The main outcome from our previous results is that the normal state has $A_n \sim T^{1/2}$ over a broad range of Matsubara frequencies, while the nonlinear gapped state respects the $T=0$ scaling and hence both $A_n$ and $\t \Delta_n$ scale like $T^{2/3}$. Therefore,
\be\label{eq:DeltaF}
F_{gapped}-F_{normal}  \approx -\frac{N T k_F}{v}\left(\gamma_0 \frac{T^{2/3}\Lambda^{1/3}}{N}-\gamma_1 \sqrt{g^2 T}\right)\,,
\ee 
with $\gamma_0$ and $\gamma_1$ order one numerical constants that depend on the particular frequency profiles, but are approximately independent on $T$.

The different temperature scaling of both regimes implies that the nonlinear gapped state is stable only at sufficiently high temperatures,
\be
\frac{T}{\Lambda} \gtrsim \left(\frac{g^2}{\Lambda} \right)^3 N^6\,,
\ee
where we are keeping implicit some $O(1)$ numerical factors. On the other hand, this state only appears for $T< T_{NL}$ as in (\ref{eq:Tsmall}). Combining both conditions, we obtain that the gapped state will then exist and be energetically preferred only if
\be
\frac{\Lambda}{g^2} \gtrsim N^3\,,
\ee
again up to $O(1)$ factors. This does not occur in general, because $\Lambda$ is a dynamical scale generated below $g^2$, that is $\Lambda< g^2$.

This discussion has incorporated only the parametric dependence on $N, T, \Lambda, g^2$, leaving some numerical constants undetermined. We have evaluated (\ref{eq:DeltaF}) on example solutions, finding agreement with this parametric analysis. 

So far we have taken into account the fermionic free energy, but the bosonic sector also contributes. The non-static modes are approximately the same in the normal and gapped state, so only the static mode contributes to the energy difference. The bosonic free energy is evaluated in detail in Appendix \ref{ap:free}, and the outcome is similar to that in the fermionic sector: the conditions that the bosonic free energy in the gapped state is smaller than in the normal state, together with $T< T_{NL}$, force the theory into a regime of parameters that is not physically realized.

To summarize, the nonlinear gapped state of Sec.~\ref{sc:gaptrunc}  and \ref{sc:num-check} is energetically disfavored compared to the normal state in the regime $T< T_{NL}$ where it exists. The dynamics and phase structure is then that of Sec.~\ref{sec:finiteT}.

\subsection{Comparing with the  ``first Matsubara law'' solution}\label{subsec:firstM}

Let us end our analysis by comparing the previous nonlinear gapped state with the ``first Matsubara law'' solution of~\cite{2016PhRvL.117o7001W, 2019PhRvB..99n4512W, 2019arXiv191201797C, Abanov_2020, Wu_2020}. These works solve gap equations that are similar to (\ref{eq:tDelta1}), but dropping $m=n$ contributions from virtual static bosons. In the procedure leading to (\ref{eq:barDelta}), this amounts to setting $N=1$ in the definition of $\bar A(\omega_n)$ there, but keeping the overall $1/N$ in the gap equation. We don't know of a field theory construction that gives this; instead, the motivation in this line of research is to understand the physics of the $N=1$ theory but artificially decreasing the strength of the pairing instability in order to reveal NFL effects. As we have stressed in previous sections, in our work the parameter $N$ is physical (see Sec.~\ref{sec:thermal}), and affects simultaneously the pairing strength and the thermal corrections. 

Despite this difference, the nonlinear gap profile we have found shares some similarities with the first Matsubara law solution. The basic reason for this is that thermal effects (neglected from the start in~\cite{2016PhRvL.117o7001W, 2019PhRvB..99n4512W, 2019arXiv191201797C, Abanov_2020, Wu_2020}) are also very small in our case when $T \ll T_{NL}$. This was explained in Sec.~\ref{sc:gaptrunc} , where the $\hat f_n$ were indeed found to be negligible if $T \ll T_{NL}$. The basic properties of the two solutions then agree at sufficiently small temperatures. Not coincidentally, these works also found a transition temperature for the first Matsubara law solution that behaves like our $T_{NL}$.

However, the dynamics and phase diagram in both approaches are quite different because, by not including thermal effects, the authors of~\cite{2016PhRvL.117o7001W, 2019PhRvB..99n4512W, 2019arXiv191201797C, Abanov_2020, Wu_2020} found that the gapped first Matsubara law state is energetically preferred. In contrast, this state is energetically disfavored in our setup, precisely because thermal terms have a different parametric scaling with temperature, see (\ref{eq:DeltaF}). Superconductivity in our case only occurs for $N<N_{cr}$ and is driven by the linearized solution of Sec.~\ref {sec:finiteT}. The details about how the gapped state disappears as $T \to T_{NL}$ are also different since these are sensitive to thermal static contributions as in Fig.~\ref{fig:contour1}.

\section{Conclusions}\label{sec:concl}

In this work we have determined the consequences from virtual static bosons on superconductivity in NFLs. Motivated by different models, we identified a parameter $N$ that measures the ratio between self-energy and gap renormalizations. Thermal contributions from such static modes cancel only when $N=1$, while for any $N>1$ they are nonzero, violate quantum-critical scaling and dominate at small temperature. We performed a detailed analysis for a Fermi surface coupled to massless Landau damped bosons in $d=2$ space dimensions, where $N$ comes from a global $SU(N)$ symmetry. For $N>N_{cr} \sim 8$, the normal state is stable under superconductivity, leading to a naked quantum critical point at $T=0$, which features critical pairing fluctuations. An infinite order transition to superconductivity appears as $N \to N_{cr}$, with the critical temperature $T_c$ and the gap $\Delta_0$ displaying BKT scaling. The superconducting phase arises then for $N< N_{cr}$ and $T<T_c$. Superconductivity exhibits strong non-Fermi liquid deviations from the usual BCS results; quantum corrections lead to an exponential suppression of $T_c$ and $\Delta_0$, and thermal effects give rise to a parametric hierarchy $T_c \ll \Delta_0$. A summary of the phase diagram can be found in Fig.~\ref{fig:phase-diag} above.

Let us suggest some directions of future research that are motivated by these results. We analyzed the role of thermal effects from static modes in a specific class of NFLs, but our basic conclusions regarding their effects when $N>1$ are likely more general. It would then be interesting to revisit other models of NFLs which feature an $N>1$, and include exchange of static modes at finite temperature. We suggest that this leads to a distinction between NFLs where superconductivity is enhanced ($N=1$) and others where it is suppressed ($N>1$), even with the same type of coupling to bosonic modes. This adds to the distinction between models with order parameters and models with emergent $U(1)$ gauge fields found in~\cite{Metlitski}. We also plan to extend our results to models with more general dynamical exponent $z_f$. These quantum critical points cover quite different phenomenology and arise also from semi-holographic constructions. 

Finally, it would be interesting to further develop strong-coupling methods to cope with two-dimensional models with $N \sim 1$, which do not have a small-parameter expansion. Our predictions of the qualitative difference between $N=1$ and $N>1$ are encouraging in this direction, because we are finding that $N$ does not need to be too large in order to observe dramatic NFL effects. Models with moderate $N$ could reasonably be studied numerically in the near future, particularly given recent developments in Monte-Carlo methods~\cite{Schattner2016, 2017PNAS..114.4905L, Berg2019, liu2019itinerant, klein2020normal, Xu:2020tvb}.


\acknowledgements
We are grateful to A. Abanov and A. Chubukov for very useful discussions and detailed comments on the manuscript. We are supported by CNEA, Conicet (PIP grant 11220150100299), ANPCyT (PICT 2018-2517) and UNCuyo.


\appendix

\section{Schwinger-Dyson equations}\label{ap:SD}

This Appendix presents the derivation of the Schwinger-Dyson equations, together with approximate expressions used in the main text.

\subsection{Notation and conventions}

In the Nambu-Gorkov basis for the fermionic fields we have
\be
\Psi_i = \left(\begin{array}{c}\psi_i(p) \\ \psi_i(-p)^\dagger\end{array}\right)\,,
\ee 
where we are collectivelly denoting $p=(\omega, \vec p)$ and $i$ denotes the $SU(N)$ flavor index. The term in the action accounting for the gap coupling is then
\be
\Psi^T_i(p)\tilde\Delta^M_{ij}\Psi_j(-p') \,, 
\ee 
with the pairing gap matrix defined as
\be
\tilde\Delta^M=\left(\begin{array}{cc}
\tilde\Delta_{ij}(p-p') & 0 \\
0 & \tilde\Delta^\dagger_{ij}(p-p') 
\end{array}\right)\,.
\ee

The Green's function reads
\be
{\cal G}_{ij}= \langle\Psi_i(p) \Psi_j(p)^\dagger\rangle =
\left(\begin{array}{cc}G_{ij}(p) & \tilde G_{ij}(p) \\ 
-\tilde G_{ij}(p)^* & G_{ij}(p)^*\end{array}\right)\,,
\label{genGreen}
\ee
where
\bea
G_{ij}(p)&=& \delta_{ij}\frac{\varepsilon_p+i A(p)}{\varepsilon_p^2+A(p)^2+|\tilde\Delta(p)|^2}\,,\label{eq:norm-prop}\\
\tilde G_{ij}(p)&=& \frac{\tilde\Delta_{ij}(p)}{\varepsilon_p^2+A(p)^2+|\tilde\Delta(p)|^2}\,,\label{eq:anom-prop}
\eea 
and we recall that $A(p)=\omega+\Sigma(p)$. 

Throughout this work, we will consider the following simple symmetry breaking pattern for the pairing vertex,
\be\label{eq:pattern}
\tilde\Delta_{ij}(p)= \tilde\Delta(p) J_{ij} \,, \; \; 
J_{ij}= \left(\begin{array}{cc}
0 & {\mathbb I}_{N/2} \\
-{\mathbb I}_{N/2} & 0
\end{array}\right)\,.
\ee
In addition, for the bosonic propagator we have
\be
D^{-1}_{ij,kl}(\Omega,q)=\delta_{il}\delta_{jk}\left(q^2+M_D^2\frac{|\Omega|}{q}\right)+\Pi_{ij,kl}(\Omega,q)\,.
\ee  
The leading contributions to the bosonic self-energy are of the form $\Pi_{ij,kl}\sim \delta_{il}\delta_{jk}$, so we can write
\bea
D^{-1}_{ij,kl}(\Omega,q)&=& \delta_{il}\delta_{jk} D^{-1}(\Omega,q)\,, \nonumber\\
D^{-1}(\Omega,q)&=&q^2+M_D^2\frac{|\Omega|}{q}+\Pi(\Omega,q)\,.\label{eq:bos-prop}
\eea

\subsection{Schwinger-Dyson-Eliashberg equations}

As shown in \cite{Damia:2020yiu}, for a vanishing gap function, self-interactions of static (zero frequency) bosonic modes have to be accounted for in order to get a self-consistent solution free of IR singularities. For higher frequencies, this interaction becomes irrelevant due to the $z_b=3$ scaling induced by Landau damping and we will ignore it. The effective action for the static mode reads
\begin{widetext}
\be
S_{\Omega_n=0}=\int \frac{d^2q}{(2\pi)^2} \frac{1}{2}\,\tr\, (\t \phi_q q^2 \t \phi_{-q} ) 
+ \frac{\lambda_\phi T}{8 N} \int \prod_{i=1}^3  \frac{d^2q_i}{(2\pi)^2}\,\tr(\t \phi_{q_1} \t \phi_{q_2}\t \phi_{q_3}\t \phi_{-q_1-q_2-q_3})\nonumber\,,
\ee
with $\tilde\phi(q)\equiv T^{1/2}\phi(\Omega_n=0,q)$. A scaling analysis shows that the coupling $\lambda_\phi T$  becomes relevant at low energies and momenta $q<(2 \pi T M_D^2)^{1/3}$. The cutoff here is set by the gap to higher Matsubara modes. For a more detailed treatment we refer the reader to \cite{Damia:2020yiu}.

Working out the contractions over flavor indices and keeping the leading terms, the self-consistent equations ignoring vertex corrections are 
\bea
\Pi(\Omega_n,q)&=& \frac{g^2T}{ N}\sum_m\int \frac{d^2p}{(2\pi)^2} \, G(p,\omega_n)  G(p+q,\Omega_m+\omega_n) +\delta_{n,0} \lambda_\phi T \int\frac{d^2 p}{(2\pi)^2} D(p,0) \label{eq:Pi-SD0}\\ 
i\Sigma(p,\omega_n)&=& g^2 T \sum_m\int\frac{d^2q}{(2\pi)^2}G(q,\omega_m)D(p-q,\omega_n-\omega_m) \label{eq:Sigma-SD0} \\
\tilde\Delta(p,\omega_n)&=& \frac{g^2 T}{ N} \sum_m\int\frac{d^2q}{(2\pi)^2}\tilde G(q,\omega_m)D(p-q,\omega_n-\omega_m) \,.\label{eq:Delta-SD0}
\eea
Recall that $\t G$ was defined in (\ref{eq:anom-prop}). We have neglected contributions from $\t G$ to the bosonic self-energy equation \eqref{eq:Pi-SD} because they are of order $N^{-2}$. Index contractions using the pattern (\ref{eq:pattern}) give the $1/N$ factor in \eqref{eq:Delta-SD}. We do not neglect this $1/N$ effect because it is the leading contribution to the gap; moreover, if the gap develops, it is always a relevant deformation to the critical point and hence becomes important at long distance.
The above equations can be alternatively obtained by minimizing the Luttinger-Ward free energy functional as shown in Appendix \ref{ap:free}.

Finally, plugging \eqref{eq:norm-prop}, \eqref{eq:anom-prop} and \eqref{eq:bos-prop} we get
\bea
\Pi(\Omega_m ,q) &=& \frac{g^2}{N} T \sum_n\,\int \frac{d^2 p}{(2\pi)^2}\,\frac{i A(\omega_n) +\varepsilon_p}{A(\omega_n)^2+|\tilde \Delta(\omega_n)|^2+\varepsilon_p^2}\frac{i A(\omega_n+\Omega_m) +\varepsilon_{p+q}}{A(\omega_n+\Omega_m)^2+|\tilde \Delta(\omega_n+\Omega_m)|^2+\varepsilon_{p+q}^2} \label{eq:Pi-SD}\\
&\,\,\,&  + \,\, \delta_{n,0} \lambda_\phi T \int\frac{d^2 p}{(2\pi)^2} \frac{1}{p^2+\Pi(0,p)}\,,\nonumber\\
i \Sigma(\omega_n) &=& g^2 T \sum_m \int \frac{d^2q}{(2\pi)^2} \frac{1}{q^2+M_D^2\frac{|\omega_n-\omega_m|}{q}+\Pi(\omega_m-\omega_n,q)}\,\frac{i A(\omega_m)+\varepsilon_{p+q}}{A(\omega_m)^2+|\tilde \Delta(\omega_m)|^2+\varepsilon_{p+q}^2}\label{eq:Sigma-SD} \\
\tilde \Delta(\omega_n) &=& \frac{g^2}{N} T \sum_m \int \frac{d^2q}{(2\pi)^2} \frac{1}{q^2+M_D^2\frac{|\omega_n-\omega_m|}{q}+\Pi(\omega_m-\omega_n,q)}\,\frac{\tilde \Delta(\omega_m)}{A(\omega_m)^2+|\tilde \Delta(\omega_m)|^2+\varepsilon_{p+q}^2}\,. \label{eq:Delta-SD}
\eea
We have already made explicit our assumption that the fermion self-energy and gap only depend on the frequency, as explained in the main text.

\subsection{Bosonic mass}\label{subsec:mb}

Let us first focus on the equation for the bosonic self-energy \eqref{eq:Pi-SD}, which we rewrite here again after expanding the fermionic energy near the Fermi surface ($\varepsilon_p\approx v p_\perp$) 
\bea
\Pi(\Omega_n,q)=&  &\frac{g^2 k_F T}{N}\sum_m \int \frac{dp_\perp d\theta}{(2\pi)^2} \frac{(iA_m+vp_\perp)(iA_{n+m}+v(p_\perp+q\cos\theta))}{(A_m^2+\tilde\Delta_m^2+(vp_\perp)^2)(A_{n+m}^2+\tilde\Delta_{n+m}^2+v^2(p_\perp+q\cos\theta)^2)}\nonumber\\
&+&\delta_{n,0} \lambda_\phi T \int\frac{d^2 p}{(2\pi)^2} \frac{1}{p^2+\Pi(0,p)}\,, \label{eq:Pi-SD2}
\eea
For non-zero frequency modes, the gap contribution is negligible and the remaining integral leads to a term of the same form as the UV Landau damping in (\ref{eq:bos-prop}), but suppressed by $1/N$. This can be neglected.

For the static modes, the situation is more subtle. At high enough temperatures we expect to have a disordered phase, so let us start by assuming a vanishing gap, {\it i.e.} $\{\tilde \Delta_n\}=\{0\}$. As described in \cite{Damia:2020yiu}, in that case the contribution from the first line in (\ref{eq:Pi-SD2}) vanishes for zero external frequency. So the Schwinger-Dyson equation gives a self-consistent equation for a boson thermal mass,
\be
\{\tilde \Delta_n\}=\{0\} \,\, \Rightarrow \,\, \Pi(0,q) \approx m^2_b =  \frac{\lambda_\phi T}{2\pi} \int \frac{p\,d p}{p^2+m^2_b}\,,
\ee
with approximate solution
\be
m_b^2 \approx \frac{\lambda_\phi T}{4\pi}\,\,\log\left(4\pi \frac{(2\pi T M_D^2)^{2/3}}{\lambda_\phi T} \right)\,.
\label{eq:thermal-mass}
\ee
The inclusion of the thermal mass \eqref{eq:thermal-mass} leads to the normal state described in \cite{Damia:2020yiu} and reviewed in Sec. \ref{subsec:normal}.

However, at low temperatures, a solution with non-trivial gap might develop, so we need to consider that case as well. For simplicity, we assume that in such a case, the first line in \eqref{eq:Pi-SD2} dominates, such that we can neglect the effect of static mode self-interactions. This assumption is shown to be consistently satisfied by the solutions found in Sec.~\ref{sc:non-linear}. In this case,
\bea
\{\tilde \Delta_n\}\neq \{0\} \, \Rightarrow \,\Pi(0,q)&\approx& 
 \frac{g^2 k_F T}{(2\pi)^2}\sum_n \int dp_\perp d\theta \frac{(iA_n+vp_\perp)(iA_n+v(p_\perp+q\cos\theta))}{(A_n^2+\tilde\Delta_n^2+(vp_\perp)^2)(A_n^2+\tilde\Delta_n^2+v^2(p_\perp+q\cos\theta)^2)}\,, \\
&=&\frac{g^2 k_F T}{4 v N}\sum_n\frac{\tilde\Delta_n^2}{(A_n^2+\tilde\Delta^2_n)\sqrt{A_n^2+\tilde\Delta^2_n+(vq/2)^2}}\,,
\eea
where, in going to the second line,we performed the momentum integration. As we are interested in the limit of small momentum, we take $q\to 0$ in the last expression above, thus obtaining
\be
\Pi(0,q)\approx m^2_{\Delta} = \frac{g^2 k_F T}{4 v N}\sum_n\frac{\tilde\Delta_n^2}{(A_n^2+\tilde\Delta^2_n)^{3/2}} \,.
\label{eq:gap-mass}
\ee 
(As we just discussed, for this to be self-consistent we must check that $m^2_{\Delta}\gg m_b^2$ once the corresponding solution is obtained, as we do in Sec.~\ref{sc:non-linear}.)

\subsection{Momentum integrals}

Now we turn to the remaining Schwinger-Dyson-Eliashberg equations \eqref{eq:Sigma-SD} and \eqref{eq:Delta-SD}. Plugging the propagators \eqref{eq:norm-prop}, \eqref{eq:anom-prop} and \eqref{eq:bos-prop}, and using the integral
\be
\int_0^{2\pi}d\theta \frac{a+b \cos\theta}{c^2+ (b\cos\theta)^2}= \frac{2\pi a}{|c|\sqrt{c^2+b^2}}
\ee
we obtain 
\bea
\Sigma(\omega_n) &=&  \frac{g^2 T}{2\pi}\sum_m  \frac{A(\omega_m)}{\sqrt{A(\omega_m)^2+\tilde\Delta(\omega_m)^2}} \int qdq\frac{D(i\omega_n-i\omega_m,q)}{\sqrt{A(\omega_m)^2+\tilde\Delta(\omega_m)^2+(vq)^2}}\,, \\
\tilde\Delta(\omega_n) &=&\frac 1 N \frac{g^2 T}{2\pi }\sum_m  \frac{\tilde\Delta(\omega_m)}{\sqrt{A(\omega_m)^2+\tilde\Delta(\omega_m)^2}} \int qdq\frac{D(i\omega_n-i\omega_m,q)}{\sqrt{A(\omega_m)^2+\tilde\Delta(\omega_m)^2+(vq)^2}}\,.
\eea

For the terms $m \neq n$ in the Matsubara sum, the boson propagator has negligible quantum corrections, from the discussion around (\ref{eq:Pi-SD2}). So splitting the sum, we write more explicitly
\bea
\Sigma(\omega_n) &=&  \frac{g^2 T}{2\pi}\left\lbrace\sum_{m\neq n}  \frac{A(\omega_m)}{\sqrt{A(\omega_m)^2+\tilde\Delta(\omega_m)^2}} \int \frac{q \, dq}{q^2+M_D^2\frac{|\omega_n-\omega_m|}{q}}\frac{1}{\sqrt{A(\omega_m)^2+\tilde\Delta(\omega_m)^2+(vq)^2}}\right.\,, \label{eq:Sigma-SD2}\\
&\,\,& \,\,\, \left. + \,\, \frac{A(\omega_n)}{\sqrt{A(\omega_n)^2+\tilde\Delta(\omega_n)^2}} \int \frac{q \, dq}{q^2+\Pi(0,q)}\frac{1}{\sqrt{A(\omega_n)^2+\tilde\Delta(\omega_n)^2+(vq)^2}}\right\rbrace \,, \nonumber
\eea
\bea
\tilde\Delta(\omega_n) &=& \frac{1}{N} \frac{g^2 T}{2\pi}\left\lbrace\sum_{m\neq n}  \frac{\tilde\Delta(\omega_m)}{\sqrt{A(\omega_m)^2+\tilde\Delta(\omega_m)^2}} \int \frac{q \, dq}{q^2+M_D^2\frac{|\omega_n-\omega_m|}{q}}\frac{1}{\sqrt{A(\omega_m)^2+\tilde\Delta(\omega_m)^2+(vq)^2}}\right.\,, \label{eq:Delta-SD2} \\
&\,\,& \,\,\, \left. + \,\, \frac{\tilde\Delta(\omega_n)}{\sqrt{A(\omega_n)^2+\tilde\Delta(\omega_n)^2}} \int \frac{q \, dq}{q^2+\Pi(0,q)}\frac{1}{\sqrt{A(\omega_n)^2+\tilde\Delta(\omega_n)^2+(vq)^2}}\right\rbrace \,,
\nonumber
\eea
where $\Pi(0,q)$ is given by either \eqref{eq:thermal-mass} for $\{\tilde\Delta_n\}=\{0\}$ or \eqref{eq:gap-mass} for $\{\tilde\Delta_n\}\neq\{0\}$.

Performing the remaining momentum integrals obtains
\bea\label{eq:SD-dim}
\hat A_n &=&\hat \omega_n + \hat f_n \frac{\hat A_n}{\sqrt{\hat A_n^2 + \hat \Delta_n^2}}+ \xi \sum_{m \neq n} \frac{1}{ |m -n|^{1/3}}\frac{\hat A_m}{\sqrt{\hat A_m^2 + \hat \Delta_m^2}} \nonumber\\
\hat \Delta_n &=&\frac{1}{N} \hat f_n \frac{\hat \Delta_n }{\sqrt{\hat A_n^2 + \hat \Delta_n^2}}+\frac{\xi}{N} \sum_{m \neq n}\frac{1}{ |m -n|^{1/3}}\frac{\hat \Delta_m }{\sqrt{\hat A_m^2 + \hat \Delta_m^2}} \,,
\eea
We have found it convenient to use the dimensionless quantities (\ref{eq:dimensionless}) and (\ref{eq:xidef}); in particular, $\hat A_n=\hat\omega_n+\hat\Sigma_n$. Furthermore,
\be
\hat f_n \equiv \frac{g^2 }{2\pi^3 T} \frac{\cosh^{-1}\left(\frac{1}{v}\sqrt{\frac{\hat A_n^2+ \hat \Delta_n^2}{\hat m^2}} \right)}{\sqrt{\hat A_n^2+ \hat \Delta_n^2 - v^2 \hat m^2}}\label{eq:fn}\,.
\ee

By considering parity properties
\be
\omega_{-n}=-\omega_{n-1} \, , \, A_{-n}=-A_{n-1} \, , \, \t \Delta_{-n}=\pm \t \Delta_{n-1}
\ee
we can rewrite the equations as sums over positive modes 
\bea\label{eq:SD-dim2}
\hat A_n &=&\hat \omega_n + \hat f_n \frac{\hat A_n}{\sqrt{\hat A_n^2 + \hat \Delta_n^2}}+ \xi \sum_{m=0}^\infty K^-_{nm} \frac{\hat A_m}{\sqrt{\hat A_m^2 + \hat \Delta_m^2}} \nonumber\\
\hat \Delta_n &=&\frac{1}{N} \hat f_n \frac{\hat \Delta_n }{\sqrt{\hat A_n^2 + \hat \Delta_n^2}}+\frac{1}{N}\xi \sum_{m=0}^\infty K^\pm_{nm} \frac{\hat \Delta_m }{\sqrt{\hat A_m^2 + \hat \Delta_m^2}} \,.
\eea
where the convolution kernel reads
\be
K^{\pm}_{nm}=\frac{1- \delta_{mn}}{|m-n|^{1/3}}\pm \frac{1}{|m+n+1|^{1/3}}\label{eq:kernel}\,.
\ee
In this work, we will focus on even solutions for $\hat\Delta_n$, which dominate because the kernel $K^+$ amounts to a larger contribution than its odd counterpart.

\section{Free Energy}\label{ap:free}

Following the Luttinger-Ward formalism \cite{LuttingerWard:1960, Benlagra:2011, ChubukovHaslinger:2003}, the free energy is $F=F_f+F_b$ with fermionic and bosonic contributions
\bea
F_f &=& -N T \sum_p \log (\epsilon_p^2+A^2+\tilde\Delta^2) +i \Sigma (G-G^*)-2\tilde\Delta \tilde G \nonumber \\
&\,\,&+ \frac{N T^2 g^2}{2} \sum_{p,p'} G(p) D(p-p') G(p')+G(p)^* D(p-p') G(p')^* -\frac{2}{N} \tilde G(p) D(p-p') \tilde G(p') \\
F_b&=&T N^2\sum_p \left[\log(D^{-1}) -\Pi D\right]+\frac{\lambda_\phi T^2 N^2}{2}\sum_{p,p'} D(p)D(p') \nonumber
\eea
where we have already performed the $SU(N)$ index contractions and assumed a real gap profile. It is direct to check that, upon variation with respect to $\Sigma$, $\tilde\Delta$ and $\Pi$, the above expression gives the correct quantum equations of motion at leading order in $N$, namely
\bea
i\Sigma(p)&=& g^2 T \sum_{p'}G(p')D(p-p') \\
\tilde\Delta(p)&=& g^2 T \sum_{p'}\tilde G(p')D(p-p') \\
\Pi(q) &=& \frac{g^2 T}{N}\sum_{p'}G(p')G(p+p')+\lambda_\phi T \sum_{p'} D(p')\label{eq:FEPi}
\eea
In the normal state, the only contribution to $\Pi(q)$ at $\Omega=0$ comes from the self-interaction term proportional to $\lambda_\phi$. On the other hand, for the gapped state solution, both the fermion bubble and the self-interaction contribute. However, by the considerations made at the end of section \ref{sc:gaptrunc}, we can neglect the second term in \eqref{eq:FEPi} for the range of temperatures for which this solution exists. 

We want to compare the free energies of the normal and gapped states, evaluated on the equations of motion. In this case, the fermion contribution takes the form
\bea
F_f &= & -\frac{N T}{2} \sum_p (2\log (\epsilon_p^2+A^2+\tilde\Delta^2) +i \Sigma (G-G^*)-2\tilde\Delta \tilde G )\\
&=&-\frac{NT k_F}{v}\sum_{n=0}^{\infty}\left( \sqrt{A_n^2+\tilde\Delta_n^2}+\frac{|\omega_n| |A_n|}{\sqrt{A^2_n+\tilde\Delta_n^2}} \right)
\eea  
where, in going to the second line, we performed the momentum integration and subtracted the logarithmically UV divergent piece (which cancels in comparing both states).    

For the bosonic piece, only the contribution of the zero mode distinguishes between the normal and gapped states. Indeed, higher frequency contributions are determined by $z=3$ scaling, being thus the same in both solutions. In addition, when evaluated in the normal state, the zero mode free energy gets a further contribution from the quartic potential. All in all we obtain 
\bea
F_{normal} &=& 
 \frac{N^2 T}{4\pi}\left(m^2_b+\frac12 m^2_b \log{\frac{\Lambda_{UV}^2}{m^2_b}} \right) + {\rm high \, freq} \,, \\
 F_{gapped} 
&=& \frac{N^2 T}{4\pi}m^2_\Delta + {\rm high \, freq} \,,
\eea  
where $\Lambda_{UV}$ is a UV cutoff for the momentum integral of the zero modes which might be suitably choosen for the case of interestest. In the expressions above, we are also neglecting a universal divergent piece which anyway cancels out when computing the condensation energy.

It is then clear that the bosonic condensation energy becomes proportional to the difference of the corresponding zero mode masses
\be
F_{gapped}-F_{normal}\sim m^2_{\Delta}-m^2_b \,.
\ee
Moreover, in section \ref{sc:non-linear} it has been shown that, as long as the gapped solution exists, that is $T<T_{NL}$, then the inequality $m^2_\Delta>m^2_b$ holds ({\it c.f.} equations \eqref{eq:mratio} and \eqref{eq:lambdaphi-cond}). So we have
\be
F_{gapped}>F_{normal} \quad , \quad T<T_{NL} \,.
\ee
Asking for the gapped state to be energetically favoured then pushes us away from the region of validity of the solution, thus leading to the same conclusion as presented in the main text for the fermionic piece.

\end{widetext}

\bibliography{NFL}{}
\bibliographystyle{utphys}
\end{document}